\documentclass[preprint,showpacs,preprintnumbers,amsmath,amssymb]{revtex4}

\usepackage{graphicx}% Include figure files
\usepackage{dcolumn}% Align table columns on decimal point
\usepackage{bm}% bold math
\usepackage{feynmp}

\begin{document}

\preprint{}

\title{Quantum Phase Diagram of Bosons in Optical Lattices}% Force line breaks with \\

\author{F. E. A. dos Santos}
 \email{santos@physik.fu-berlin.de}
\affiliation{%
Institut f\"{u}r Theoretische Physik, Freie Universit\"{a}t Berlin, Arnimallee 14, 14195 Berlin, Germany}%

\author{A. Pelster}
\email{axel.pelster@uni-duisburg-essen.de}
\affiliation{
Fachbereich Physik, Universit\"{a}t Duisburg-Essen, Campus Duisburg, Lotharstrasse 1, 47048 Duisburg, Germany
}%

\date{\today}% It is always \today, today,
             %  but any date may be explicitly specified

\begin{abstract}
We work out two different analytical methods for calculating the boundary of the Mott-insulator-superfluid (MI-SF) quantum phase transition for scalar bosons in cubic optical lattices of arbitrary dimension at zero temperature which improve upon the seminal mean-field result.
The first one is a variational method, which is inspired by variational perturbation theory, whereas the second one is based on the field-theoretic concept of effective potential. Within both analytical approaches we achieve a considerable improvement of the location of the MI-SF quantum phase transition for the first Mott lobe in excellent agreement with recent numerical results from Quantum Monte-Carlo simulations in two and three dimensions. Thus, our analytical results for the whole quantum phase diagram can be regarded as being essentially exact for all practical purposes.

% as well as Dynamical Renormalization Group calculations for 1 dimension.

%Such quantities are obtained from the free energy of Bose-Hubbard Hamiltonian with extra global-current %terms. The free energy in our zero temperature case is given by the ground-state energy which here is %calculated perturbatively up to second order in the hopping parameter $t$ by using two diagrammatic %methods. Here this expansion is applied in two different methods.
%In section III we work out our first method where we perform systematic variational %corrections to the previous mean-field calculations by considering a different %interpretation for the order parameter $\psi$. Unlike in mean-field theory, where %$\psi$ is determined from self-consistence relations, here $\psi$ must be seen as a %variational parameter like in {\it variational perturbation theory}. In this new %approach $\psi$ can still be found from extremalizing of the grand-canonical free %energy, however for the variational approach the self-consistence relations no longer %apply.

%Afterwards we present our second approach in section IV which is based a standard %field-theoretical method. Here we add spatially and temporally global source terms to %the usual Bose-Hubbard Hamiltonian in order to break the global $U(1)$ symmetry. 
%The Legendre transformation of the grand-canonical free energy obtained from this %modified Hamiltonian defines an effective potential which is calculated up to second %order in the hopping parameter $t$.

\end{abstract}

% PACS, the Physics and Astronomy
                             % Classification Scheme.
%\keywords{Suggested keywords}%Use showkeys class option if keyword
                              %display desired
\maketitle

\begin{fmffile}{graph}
\fmfstraight
\section{\label{sec:level1} Introduction}

During the last few years, experiments on trapped dilute ultracold quantum gases led to the observation of many new novel properties of these quantum systems \cite{boseold7,contex,boseold8,grimm}. Among these systems are ultracold bosonic gases trapped in the periodic potential of optical lattices. These led to a whole plethora of experimental possibilities on many-particle quantum physics as they represent model systems for solid-state physics with a yet unprecedented level of control \cite{collapse,mandel,gerbierpra,gerbier,FW06,zurich,OOW06,lewenstein,inguscio}. Optical lattices can be formed by using electromagnetic standing waves orthogonally aligned to each other, with their crossing point positioned at the center of a Bose-Einstein condensate. In this way, they generate a periodic potential where the atoms can move from one lattice site to the next due to the quantum mechanical tunnelling effect \cite{grimm}. Because of the periodicity of the optical lattice, the single-particle energy spectrum has a band structure. At low enough temperatures, the thermal fluctuations are too weak to excite the atoms beyond the lowest band. Thus, this system can be well described by the Bose-Hubbard model \cite{fischer,jaksch,sachdev,zoller}.

Such bosonic gases in optical lattices can exist in two different phases which can be chosen by tuning the depth of the potential wells generated by the optical waves. The Bose-Hubbard model states that the existence of these two phases is determined by the balance between the atom-atom on-site interaction $U$ and the hopping amplitude $t$. When the on-site interaction is small compared to the hopping amplitude,
the ground state is superfluid (SF), as the bosons are delocalized and phase coherent over the whole lattice. In the opposite limit, where the on-site interaction dominates over the hopping term, the ground state is a Mott insulator (MI), as each boson is trapped in one of the respective potential minima. These different phases are observable, for instance, in time-of-flight absorption pictures which are taken after switching off the lattice potential. While the superfluid phase yields distinct Bragg-like interference peaks, the Mott phase is characterized by a broad diffusive interference pattern \cite{gerbierpra,gerbier}.

\begin{figure}[t]
     \centerline{\includegraphics[scale=0.8]{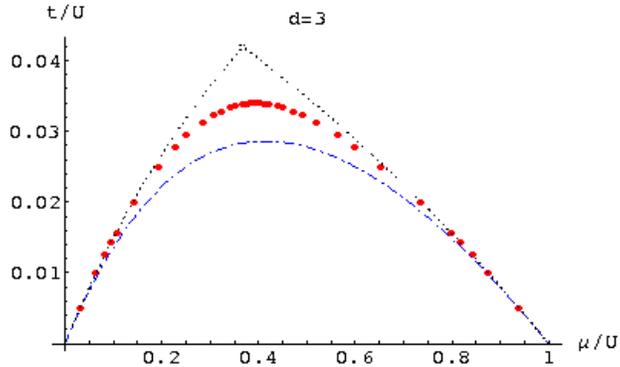}}
     \caption{(Color online) Quantum phase diagram of the first MI-SF lobe at $T=0$ for the three-dimensional case.
      Dot-dashed blue line is the mean-field result \cite{fischer}, dotted black line is from the third-order strong-coupling expansion \cite{monien}, and red dots are recent high-precision Monte-Carlo data \cite{montecarlo3}.}
      \label{fig0}
\end{figure}

A common approximation for calculating the SF-MI phase boundary uses a mean-field theory where the non-local Bose-Hubbard Hamiltonian is substituted by an effective local one \cite{fischer}. An alternative way to obtain the SF-MI phase boundary at $T=0$ is based on a strong-coupling expansion as is worked out in detail in Ref. \cite{monien}. However, comparing both analytical methods with recent high-precision Monte-Carlo data \cite{montecarlo3}, as shown in Fig.~\ref{fig0} for the three-dimensional case, we observe that the mean-field theory underestimates the location of the quantum phase transition, while the strong-coupling approach overestimates it. Thus, in view of a more quantitative comparison with experimental results, it becomes indispensable to further develop analytical approximation methods (see, for instance, 
Refs.~\cite{schroll,vollhardt,ziegler1,ziegler2,ziegler3}). In particular, obtaining accurate analytical results for the phase boundary at arbitrary dimension $d$ and lobe number $n$ would yield new insight beyond the purely numerical data 
provided by Monte-Carlo simulation. Furthermore, it would be favorable to develop analytical approaches which allow, at least in principle, systematic improvements to higher orders.

To this end we present here two alternative analytical methods to approximately solve the homogeneous Bose-Hubbard Hamiltonian and calculate the properties of the zero-temperature quantum phase diagram, in particular the location of the MI-SF phase boundary. Both our approaches are technically based on a systematic expansion with respect to the hopping parameter $t$. This perturbative procedure is justified as the phase boundary occurs in three dimensions for small values of $t/U$ (see Fig. \ref{fig0}). It turns out that this $t$-expansion can be worked out analytically up to higher orders which contain non-trivial information about the dimension $d$, and, therefore, improve considerably the mean-field result for lower dimensions. Note that our approaches essentially differ from the promising, recently developed Bosonic Dynamical Mean-Field Theory (BDMFT) which is based on a systematic $1/d$ expansion \cite{vollhardt}. Although BDMFT
has the virtue of being nonperturbative in the system parameters $t$ and $U$, its self-consistency equations can only be solved numerically.

In Section II we briefly review the usual mean-field theory which already gives a good qualitative description of the
MI-SF quantum phase transition. In Section III we work out our first method where we perform systematic variational corrections to the previous mean-field calculations by considering a different interpretation of the order parameter $\psi$. Unlike in mean-field theory, where $\psi$ is determined from self-consistency relations, here $\psi$ is regarded as a variational parameter like in {\it variational perturbation theory} \cite{kleinert2, kleinert1}. In this new approach, the order parameter $\psi$ is found by extremizing the grand-canonical free energy, although self-consistency relations no longer apply in general.
Afterwards, we present our second approach in Section IV which is based on a standard field-theoretical method. Here we add spatially and temporally global source terms to the usual Bose-Hubbard Hamiltonian in order to break the global $U(1)$ symmetry. 
The Legendre transformation of the grand-canonical free energy obtained from this modified Hamiltonian defines an effective potential which is used for determining the MI-SF phase-boundary. Both analytical approaches are calculated up to second order in the hopping parameter $t$ in Section V, while detailed technical calculations are relegated to the Appendices. Finally, we present in Section VI the resulting improved quantum phase diagrams and compare them with previous findings. In particular, we show that our analytical results have an accuracy for the first MI-SF lobe  in two and three dimensions which is comparable with Monte-Carlo data. Therefore, we are confident that our analytical results are essentially exact for any Mott lobe in more than one dimension.

\section{Mean-Field Theory}
In the present work we deal with a system of spinless bosons in a homogeneous infinite cubic lattice of
arbitrary dimension $d$, which can be well described by the Bose-Hubbard Hamiltonian \cite{fischer,jaksch,sachdev,zoller}
\begin{equation}
\label{Hubbard}
\hat{H}_{\rm BH} = -t\sum_{\langle i, j\rangle}\hat{a}^{\dag}_{i}\hat{a}_{j} + \hat{H}_{0}
\end{equation}
with $\langle i,j \rangle$ running over the nearest neighbor sites and with the on-site Hamiltonian
\begin{equation}
\label{lcl}
 \hat{H}_{0} = \sum_{i}\hat{H}_{i}, \qquad \hat{H}_{i} = \frac{U}{2}\hat{n}_{i}\left(\hat{n}_{i} - 1\right) 
- \mu\hat{n}_{i} .
\end{equation}
Here $\hat{n}_{i} = \hat{a}^{\dag}_{i}\hat{a}_{i}$ represents the number operator at site $i$, and $\mu$ denotes the chemical potential. Furthermore, $t$ is the hopping energy which characterizes the tunneling of an atom from one lattice site to a neighboring one. Furthermore, $U$ denotes the on-site energy which describes the strength of the interaction between
two atoms at a given lattice site. When the depth of the lattice wells is increased, the hopping energy $t$ decays exponentially
fast, whereas the on-site energy $U$ increases algebraically \cite{zwerger}.

A standard approach to approximately solve the Hamiltonian (\ref{Hubbard}) uses a mean-field ansatz \cite{fischer}. To this end, the non-local hopping term in (\ref{Hubbard}) is substituted by a sum of single-site terms, thus yielding the following mean-field Hamiltonian:
\begin{equation}
\label{mHubbard}
\hat{H}_{\rm MF} = -t (2d) \sum_i\left( \psi^{\ast}\hat{a}_i+ \psi\hat{a}^{\dag}_i - \psi^{\ast}\psi \right)  + \hat{H}_0 .
\end{equation}
The thermodynamical quantities are then calculated from the grand-canonical free energy
\begin{equation}
F_{\rm MF}(\psi^{\ast}, \psi) = -\frac{1}{\beta} \ln Z_{\rm MF}(\psi^{\ast}, \psi) ,
\end{equation}
where the grand-canonical partition function is given by 
\begin{equation}
 Z_{\rm MF}(\psi^{\ast}, \psi) = {\rm Tr}\left[e^{-\beta \hat{H}_{\rm MF}(\psi^{\ast}, \psi)}\right] .
\end{equation}
The additional complex parameters $\psi$ and $\psi^{\ast}$, which effectively describe the influence of the neighboring sites, have to be self-consistently determined according to $\psi^{\ast} = \langle\hat{a}^ {\dagger}_i\rangle$, $\psi = \langle\hat{a}_i\rangle$. Note that imposing these self-consistency conditions is equivalent to extremizing the grand-canonical free energy with respect to $\psi$ and $\psi^{\ast}$ :
\begin{equation}
 \begin{cases}
  {\displaystyle \frac{\partial F_{\rm MF}}{\partial\psi}} = 0 \\[2mm]
  {\displaystyle \frac{\partial F_{\rm MF}}{\partial\psi^{\ast}}} = 0
 \end{cases}
\Longrightarrow
 \begin{cases}
  \langle\hat{a}_i^{\dagger} \rangle = \psi^{\ast} \\
  \langle\hat{a}_i \rangle = \psi .
 \end{cases}
\end{equation}
Near the phase boundary the order parameter $\psi$ is small, so the grand-canonical free energy can be Taylor expanded with respect to the order parameter in form of a Landau expansion \cite{kleinert2,zinn-justin}:
\begin{equation}
\label{landau}
F_{\rm MF}(\psi^{\ast}, \psi) = N_s\left[a^{\rm MF}_0(T) + a^{\rm MF}_2(T) \vert\psi\vert^2 + a^{\rm MF}_4(T) \vert\psi\vert^4 + \cdots\right] .
\end{equation}
Here $N_s$ is the total number of lattice sites within the system. If $a_4(T) > 0$ and $a_2(T)$ changes its sign depending on the values of the respective parameters $t$, $U$, and $\mu$, then the system exhibits a second-order phase transition. The phase boundary is given by points in the parameter space where $a^{\rm MF}_2(T) =0$ is valid. At zero temperature the resulting phase boundary reads \cite{fischer}: 
\begin{equation}
\label{tcmf}
t_c^{\rm MF}=\frac{U}{2d}\left(\frac{n+1}{n-b}+\frac{n}{1-n+b}\right)^{-1} \quad,\qquad b=\frac{\mu}{U}.
\end{equation}
%
%As we will see later, all formulas derived in this paper coincides with with %(\ref{tcmf} ) due to the fact that the 
Note that this result of mean-field theory becomes exact in the limit $d\rightarrow\infty$ \cite{sachdev}.
\section{Variational Method}
As the mean-field results differ
considerably from the latest Quantum Monte-Carlo results (see Fig.~\ref{fig0}), it is necessary to develop new analytical approaches for studying bosons in optical lattices. Therefore, the objective of the present work is to generalize the mean-field approach and to calculate systematic
corrections which are due to quantum fluctuations. Our first method is based on introducing an artificial smallness parameter $\eta$ in the Bose-Hubbard Hamiltonian according to
\begin{equation}
\label{hxi}
\hat{H}\left(\eta\right) = \hat{H}_{\rm MF} + \eta \left(\hat{H}_{\rm BH}-\hat{H}_{\rm MF}\right),
\end{equation}
which can be explicitly written as:
\begin{equation}
\label{hamiltonian1}
 \hat{H} (\eta, \psi^{\ast}, \psi ) = \hat{H}_0 -t \eta \sum_{\langle i, j \rangle} \hat{a}_i^{\dagger} \hat{a}_j 
  -2d t (1 -\eta )  \sum_i \left( \psi^{\ast}\hat{a}_i 
  + \psi \hat{a}_i^{\dagger} -\vert\psi\vert^2 \right).
\end{equation}
Note the limiting cases $\eta = 0$ and $\eta =1$, where the Hamiltonian (\ref{hamiltonian1}) reduces to $\hat{H}\left(\eta =0\right)=\hat{H}_{\rm MF}$ and $\hat{H}\left(\eta =1\right)=\hat{H}_{\rm BH}$, respectively. Furthermore, the order parameter $\psi$ is treated as a variational parameter like in \textit{variational perturbation theory} (VPT) \cite{kleinert2,kleinert1}. Using (\ref{hxi}), we perform a Taylor expansion up to the $N$th order of the grand-canonical free energy in $\eta$ and obtain, as in the mean-field case, a Landau expansion for this $N$th order grand-canonical free energy:
\begin{equation}
\label{landau2}
 F^{(N)} \left( \eta, \psi^{\ast},\psi \right) = N_{\rm S} \left[ a_0^{(N)}( \eta ) + a_2^{(N)}( \eta ) \vert\psi\vert^2 + a_4^{(N)}( \eta ) \vert\psi\vert^4 + \cdots \right] .
\end{equation}
Here $a_{2p}^{(N)}$ is the truncated expansion in $\eta$ up to order $N$ of $a_{2p}(\eta)$ which is given by the expressions
\begin{equation}
\label{truncated}
 \begin{cases}
  a_2( \eta) = (2dt)^2 (1-\eta)^2 \sum_{m=0}^{\infty}(-t\eta)^m \alpha_2^{(m)}+2dt(1-\eta), \\
  a_{2p}( \eta) = (2dt)^{2p} (1-\eta)^{2p} \sum_{m=0}^{\infty}(-t\eta)^m \alpha_{2p}^{(m)} \;\;\; ;\;\;\; p\neq 1 .
 \end{cases}
\end{equation}
When we set $p=1$ and $\eta =1$, the truncated expansions reduce to 
\begin{equation}
\label{a2eta1}
\begin{cases}
 a_2^{(0)}(\eta =1) = \alpha_2^{(0)} (2d)^2 t^2 + 2d t ,\\
 a_2^{(N)}(\eta =1) = (-1)^N t^2 \left[ \alpha_2^{(N)} t^N+ \alpha_2^{(N-1)} t^{N-1} \right] \; ; \;\;\; N \geq 1 .
\end{cases}
\end{equation}
Finally, we find the phase boundary at the points where we have $a^{(N)}_2( \eta = 1) = 0$.
Thus, we conclude from (\ref{a2eta1}) that the phase boundary is given by
\begin{equation}
\label{boundary1}
\begin{cases}
\tilde{t}_c^{(0)}=-\frac{1}{(2d) \alpha_2^{(0)}}, \\
\tilde{t}_c^{(N)}=-\frac{\alpha_2^{(N-1)}}{\alpha_2^{(N)}} \; ; \;\;\; N \geq 1 .
\end{cases}
\end{equation}
Before we calculate explicitly the respective perturbative coefficients $\alpha_2^{(N)}$ in (\ref{boundary1}), we introduce in the next section our second method for determining the location of the quantum phase transition.

\section{Field-theoretic Method}
The second method developed here in order to improve the analytical results for bosons in optical lattices is not based on any mean-field theory. Instead of this, as a starting point, we consider the Bose-Hubbard Hamiltonian with additional source terms, which are spatially and temporally global:
\begin{equation}
\label{hamiltonian2}
 \hat{H}_{\rm BH}(J^{\ast}, J)= -t\sum_{\langle i, j\rangle}\hat{a}^{\dag}_{i}\hat{a}_{i} + \sum_i \left( J^{\ast} \hat{a}_i + J\hat{a}_i^{\dagger}\right) +  \hat{H}_{0}.
\end{equation}
The grand-canonical free energy is then calculated in a power series of both the hopping parameter $t$ and the 
sources $J$, $J^{\ast}$. This leads to
\begin{equation}
\label{free.energy}
F(J^{\ast}, J, t) = N_{\rm s} \left( F_0(t) + \sum_{p=1}^{\infty} c_{2p}(t) \vert J\vert^{2p} \right)
\end{equation}
with the expansion coefficients
\begin{equation}
\label{coefficients}
 c_{2p}(t) = \sum_{n=0}^{\infty} (-t)^n \alpha_{2p}^{(n)}.
\end{equation}
We observe that due to the similarities between the Hamiltonians (\ref{hamiltonian1}) and (\ref{hamiltonian2}), the coefficients $\alpha_{2p}^{(n)}$ which appear in the expression (\ref{coefficients}) are identical to the ones in Eqs.~(\ref{truncated}).

In this approach we define the order parameter $\psi$, as usual in field theory \cite{kleinert2,zinn-justin},  according to
\begin{equation}
 \psi = \langle \hat{a}_i \rangle = \frac{1}{N_{\rm s}}\frac{\partial F(J^{\ast}, J)}{\partial J^{\ast}} \;\;\; ; \;\;\;
 \psi^{\ast} = \langle\hat{a}_i^{\dagger} \rangle =  \frac{1}{N_{\rm s}}\frac{\partial F(J^{\ast}, J)}{\partial J}.
\end{equation}
A subsequent Legendre transform of the grand-canonical free energy yields the effective potential
\begin{equation}
\label{effective}
 \Gamma(\psi^{\ast}, \psi) =  F/N_{\rm s} -\psi^{\ast}J - \psi J^{\ast}.
\end{equation}
Thus, the external sources can be written as derivatives of the effective potential
\begin{equation}
 \label{sources}
 \frac{\partial \Gamma}{\partial \psi^{\ast}}= - J \;\;\; , \;\;\;
 \frac{\partial \Gamma}{\partial \psi} = - J^{\ast}.
\end{equation}
From Eq.~(\ref{sources}) we read off that the physical limit of vanishing currents is obtained by simply extremizing the effective potential with respect to $\psi$ and $\psi^{\ast}$:
\begin{equation}
 \frac{\partial \Gamma}{\partial \psi^{\ast}}= 0 \;\;\; ; \;\;\;
 \frac{\partial \Gamma}{\partial \psi} = 0 .
\end{equation}
Using (\ref{free.energy}), the effective potential (\ref{effective}) is  written as a power series of  $\vert\psi\vert^2$:
\begin{equation}
\Gamma (\psi^{\ast}, \psi, t) = F_0(t) - \frac{1}{c_2(t)}\vert\psi\vert^2 + \frac{c_4(t)}{c_2(t)^4}\vert\psi\vert^4 +\cdots .
\end{equation}
The respective coefficients can be determined as a power series in $t$. For instance, the coefficient of $\vert\psi\vert^2$ turns out to have the following hopping expansion:
\begin{equation}
 \frac{1}{c_2(t)}=\frac{1}{\alpha_2^{(0)}}\left\{ 1 + \frac{\alpha_2^{(1)}}{\alpha_2^{(0)}} t + \left[ \left( \frac{\alpha_2^{(1)}}{\alpha_2^{(0)}} \right)^2 - \frac{\alpha_2^{(2)}}{\alpha_2^{(0)}} \right] t^2 + \cdots \right\} .
\end{equation}
This expansion of $1/c_2$ in power series of $t$ is equivalent to a resummation of $c_2$ such that it has a divergency at the phase boundary. The presence of such a divergency comes from the long-range correlations of the system and is an essential feature for the occurrence of the phase transition. Thus, the quantum phase boundary is found by setting $1/{c_2(t_c)} =0$. This procedure gives us an algebraic equation in $t_c$ whose degree depends on the order of the expansion in $t$. Such an algebraic equation can have, in principle,  many real roots depending on its degree. However, only the smallest root must be considered as physical. The other roots must be discarded as they are artificially introduced within our present method whose validity is restricted to small values of $t$.
Following these criteria we find in first hopping order:
\begin{equation}
\label{f.order}
 t_c^{(1)}=-\frac{\alpha_2^{(0)}}{\alpha_2^{(1)}} .
\end{equation}
For the second order we get correspondingly:
\begin{equation}
\label{roots}
t_c^{(2)}=\frac{\overline{\alpha}_1}{2 \left( \overline{\alpha}_2 - \overline{\alpha}_1^2 \right)} + \frac{1}{2 \left( \overline{\alpha}_2 - \overline{\alpha}_1^2
\right) }  \sqrt{ \overline{\alpha}_1^2 -4\left( \overline{\alpha}_1^2 - \overline{\alpha}_2 \right)} ,
\end{equation}
where we have introduced the reduced quantities $\overline{\alpha}_1 = \alpha_2^{(1)}/\alpha_2^{(0)}$ and $\overline{\alpha}_2 = \alpha_2^{(2)}/\alpha_2^{(0)}$. Note that only the smallest root was taken into account in (\ref{roots}).

\section{Perturbative calculations}
In this paper we restrict ourselves to work out the zero-temperature limit of our two analytical approaches. Therefore, we can use the standard Rayleigh-Schr\"{o}dinger perturbation theory in order to obtain the ground-state energy corresponding to the Hamiltonians presented here. Obviously, the Hamiltonian (\ref{hamiltonian2}) is converted into the Hamiltonian (\ref{hamiltonian1}) by the following  transformation of its parameters:
\begin{eqnarray}
\label{transformations}
 t &\longrightarrow& \eta t , \nonumber \\
 J &\longrightarrow&  2 d t (\eta -1)\psi , \nonumber \\
 J^{\ast} &\longrightarrow& 2 d t (\eta -1)\psi^{\ast} , \\
 \hat{H} &\longrightarrow& \hat{H} +2 d t (1 - \eta) \vert\psi\vert^2 . \nonumber
\end{eqnarray}

This fact enables us to perform all calculations using the Hamiltonian of the field-theoretic method (\ref{hamiltonian2}) and then obtain the results corresponding to the variational method by redefining its parameters. Therefore, all we need to  calculate is the ground-state energy of the Hamiltonian (\ref{hamiltonian2}).

As the ground-state energy of (\ref{hamiltonian2}) is an extensive quantity, its calculation as a power series in the hopping parameter $t$ can be performed by applying the linked cluster method (see Appendix \ref{appA}). As can be seen in (\ref{coefficients}), the coefficients $\alpha_{2p}^{(n)}$ are the fundamental blocks for calculating all quantities which are related to the ground-state energy. In our version of the linked-cluster expansion each of the coefficients $\alpha_{2p}^{(n)}$ can be represented by a set of diagrams.
Such diagrams are composed of oriented lines linked at their ends to points which represent the vertices of the lattice. The diagrams are embedded in the lattice and, therefore, their topologies are defined by the topology of the underlying lattice. The linked-cluster theorem states that only connected diagrams contribute to extensive quantities like the ground-state energy.

As explained in detail in Appendix \ref{appB}, for a given diagram, there are lines which have both ends linked to neighbor points of the lattice (internal lines) and there are lines which are linked to a point of the lattice by only one of its ends (external lines). The internal lines are associated with the hopping of the atoms between two neighboring points and, therefore, the order in the hopping parameter $t$ of a diagram is given by the number of its internal lines. The external lines can be directed into or out of a lattice point. If the line is directed into the point it represents a creation operator acting at this point which is the coefficient of $J$ in (\ref{hamiltonian2}). In the opposite case where the line is directed out of the point it represents an annihilation operator acting at this point which is the coefficient of $J^{\ast}$ in (\ref{hamiltonian2}). Analogously to the internal lines the power of $J$ is given by the number of lines going into the diagram while the power of $J^{\ast}$ is given by the number of lines coming  out of the diagrams. As the ground-state energy of (\ref{hamiltonian2}) depends only on $J^{\ast}J$, the number of ingoing lines must be equal to the number of outgoing lines.

The location of the MI-SF phase boundary is determined by the coefficients $\alpha_2^{(n)}$ whose diagrams involve two external lines. The zeroth-order coefficient $\alpha_2^{(0)}$ corresponds to diagrams with no internal lines, thus, for topological reasons, only one diagram can contribute:

\begin{equation}
\label{aI}
\alpha_2^{(0)} = \parbox{20mm}{\begin{fmfgraph*}(60,20)
  \fmfleft{i1}
  \fmfright{o1}
  \fmf{fermion,label.side=left}{i1,v1}
  \fmf{fermion,label.side=left}{v1,o1}
  \fmfv{decor.shape=circle, decor.filled=1, decor.size=.2h}{v1}
  \end{fmfgraph*}} \qquad. 
\end{equation}
In first order the coefficient $\alpha_2^{(1)}$ corresponds to one internal line, so also there only one diagram is possible:

\begin{equation}
\label{aII}
\alpha_2^{(1)} = (2d) \; \parbox{20mm}{
\begin{fmfgraph*}(60,20)
  \fmfleft{i1}
  \fmfright{o1}
  \fmf{fermion,label.side=left}{i1,v1}
  \fmfv{decor.shape=circle, decor.filled=1, decor.size=.2h}{v1}
  \fmf{fermion, label.side=left}{v1,v2}
  \fmfv{decor.shape=circle, decor.filled=1, decor.size=.2h}{v2}
  \fmf{fermion, label.side=left}{v2,o1} 
  \end{fmfgraph*}} \;\; .
\end{equation}
Here $2d$ is the corresponding lattice number which can be explained as follows. Once the first vertex is chosen somewhere in the lattice, there are precisely $2d$ possibilities for the second vertex. The situation is more complicated for the second-order coefficient $\alpha_2^{(2)}$, as in that case two topologically different diagrams contribute:

\begin{equation}
\label{aIII}
\alpha_2^{(2)} = (2d)(2d -1) \; \parbox{20mm}{
\begin{fmfgraph*}(60,20)
  \fmfleft{i1}
  \fmfright{o1}
  \fmf{fermion,label.side=left}{i1,v1}
  \fmfv{decor.shape=circle, decor.filled=1, decor.size=.2h}{v1}
  \fmf{fermion, label.side=left}{v1,v2}
  \fmfv{decor.shape=circle, decor.filled=1, decor.size=.2h}{v2}
  \fmf{fermion, label.side=left}{v2,v3} 
  \fmfv{decor.shape=circle, decor.filled=1, decor.size=.2h}{v3}
  \fmf{fermion, label.side=left}{v3,o1} 
  \end{fmfgraph*}} \; + (2d) \; \parbox{20mm}{
  \begin{fmfgraph*}(60,40)
  \fmfleft{i1}
  \fmfright{o1}
  \fmfbottom{p1}
  \fmftop{u1}
  \fmf{fermion}{i1,v1}
  \fmfv{decor.shape=circle, decor.filled=1, decor.size=.1h}{v1}
  \fmf{fermion,tension=0,left=0.4}{v1,u1}
  \fmfv{decor.shape=circle, decor.filled=1, decor.size=.1h}{u1}
  \fmf{fermion,tension=0,left=0.4}{u1,v1}
  \fmf{fermion}{v1,o1} 
  \end{fmfgraph*}} \qquad .
\end{equation}

In Appendix \ref{appC} we develop a second kind of diagrammatics which is useful for simplifying the calculations of the coefficients (\ref{aI})--(\ref{aIII}) within time-independent perturbation theory. This explicit evaluation is relegated to Appendix \ref{appD} which yields the following expressions:

% Such diagrams should not be confused with the first kind of diagrams discussed so far %in this work.

\begin{subequations}
\label{formulas}
\begin{equation}
\label{alpha0}
\alpha_2^{(0)} =\frac{b+1}{U(b-n)(b+1-n)},
\end{equation}

\begin{equation}
\label{alpha1}
\alpha_2^{(1)} = \frac{2d (b+1)^2}{U^2(b-n)^2 (b+1-n)^2},
\end{equation}

\begin{eqnarray}
\label{alpha2}
\alpha_2^{(2)}  &=&  2d \left\{ 2d (b+1)^3 (b-2-n)(b+3-n)+n(b-n)(b+1-n)(1+n)
  \right. \nonumber  \\ && \left. \times 
  (4+3b+2n)\left[ -3-2n+2(b^2 +b-2bn+n^2) \right] \right\} \nonumber
   \\  && / \left[ U^3 (b-n-2)(b-n)^3 (b+1-n)^3 (b+3-n) \right] ,
\end{eqnarray}
\end{subequations}
where we have used again the abbreviation $b=\mu /U$.

\section{Quantum phase diagrams}
In order to calculate the quantum phase diagram by using the variational method, we have to substitute
the coefficients (\ref{formulas}) into Eqs.~(\ref{boundary1}), while for the field-theoretic method they are substituted into Eqs.~(\ref{f.order}) and (\ref{roots}).

\begin{figure}[t]
     \centerline{\includegraphics[scale=0.67]{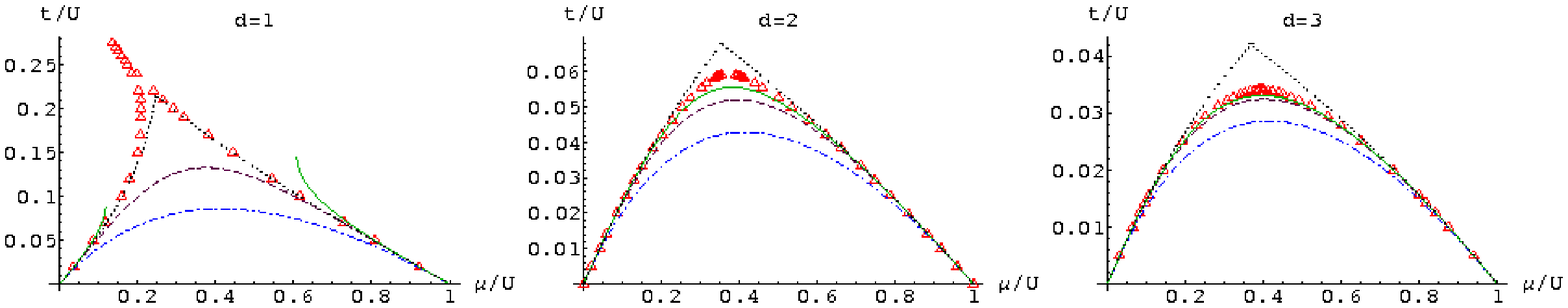}}
     \caption{(Color online) Quantum phase diagram of the first MI-SF lobe $(n=1)$ at $T=0$.
      Solid green lines are results from our field-theoretic method, dashed purple lines are results from our variational method, dot-dashed blue lines are from mean-field
      theory \cite{fischer}, dotted black lines are from the third-order strong-coupling expansion \cite{monien}, and red triangles are the numerical data. For the
      one-dimensional case the numerical data stem from Density-Matrix-Renormalization Group calculations \cite{montecarlo1}, while the data for two and three dimensions are obtained from 
      Quantum Monte-Carlo simulations \cite{montecarlo3,montecarlo2}.}
      \label{fig1}
\end{figure}

\begin{table}
\caption{\label{table}Relative deviation of our analytical findings from the Quantum Monte-Carlo data in the position of the first MI-SF lobe tip in two and three dimensions. }
\begin{ruledtabular}
\begin{tabular}{ccccc}
 &mean-field theory&strong coupling&variational method&field-theoretic method\\ \hline
 $d=2$&$28 \%$&$13 \%$ &$13 \%$&$6 \%$ \\
 $d=3$&$16 \%$&$24 \%$&$5 \%$&$3 \%$\\
\end{tabular}
\end{ruledtabular}
\end{table}

\begin{figure}[t]
     \centerline{\includegraphics[scale=0.67]{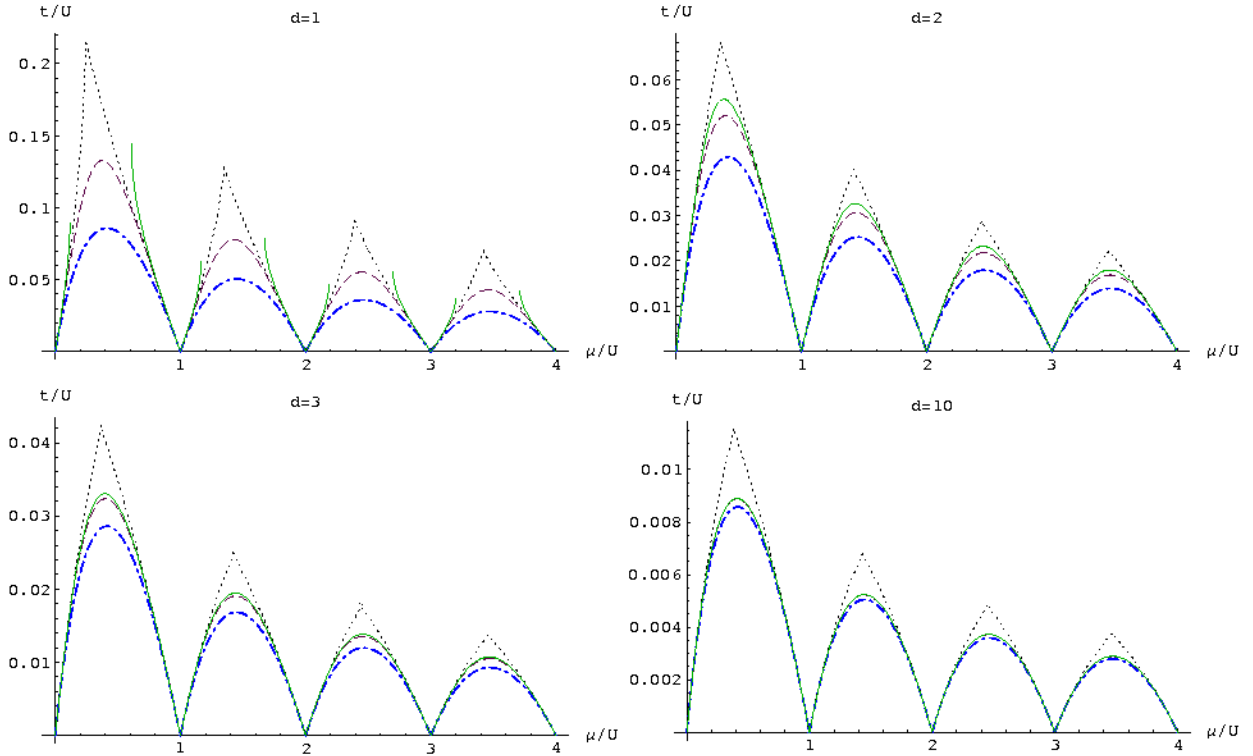}}
     \caption{(Color online) Phase diagram of MI-SF lobes at $T=0$ for different dimensions.
      Solid green line are results from our field-theoretic method, dashed purple lines are results from our variational method, dot-dashed blue lines are from mean-field
      theory, and dotted black lines are from the third-order strong coupling-expansion \cite{monien}. The non-physical $10$-dimensional case is included to 
      show that our two methods converge to the mean-field theory \cite{fischer} in the limit $d\rightarrow\infty$, as is expected on general grounds \cite{sachdev}.}
      \label{fig2}
\end{figure}

In the variational method, the  zeroth-order phase boundary reads
\begin{equation}
\label{z.th}
 \tilde{t}_c^{(0)}=-\frac{1}{(2d) \alpha_2^{(0)}} ,
\end{equation}
whereas the field-theoretic method yields in lowest order

\begin{equation}
\label{f.th}
 t_c^{(1)}=-\frac{\alpha_2^{(0)}}{\alpha_2^{(1)}} .
\end{equation}
By applying the identities (\ref{star}) we see that Eqs.~(\ref{z.th}) and (\ref{f.th}) are actually one and the same expression, which reduces with the help of (\ref{alpha0}) to the mean-field result (\ref{tcmf}).

The first non-trivial quantum correction to the quantum phase boundary in both approaches becomes available just at the second-order level, where the formulas (\ref{boundary1}) and (\ref{roots}) have to be applied together with (\ref{formulas}). In Fig.~\ref{fig1} we compare, for the first MI-SF lobe, our two methods with the mean-field phase boundary \cite{fischer}, with the third-order strong-coupling expansion \cite{monien}, and with recent numerical data obtained by using the  Density-Matrix Renormalization Group technique \cite{monien2} for one dimension as well as the Quantum Monte-Carlo \cite{montecarlo3,montecarlo2} simulations for two and three dimensions. In Table \ref{table} we compare our two analytical methods quantitatively with the Quantum Monte-Carlo data for the position of the first MI-SF lobe tip for two and three dimensions.  Thus, we can conclude from Fig. \ref{fig1} and Table \ref{table} that our analytical methods have provided very accurate results for the first MI-SF lobe in comparison with Monte-Carlo data. Therefore, we expect that they are essentially exact for any MI-SF lobe in more than one dimension where no Monte-Carlo simulations have been yet performed systematically. In addition, we read off from Fig.~\ref{fig1} and Table \ref{table} that our field-theoretic method turns out to give better results for the MI-SF phase boundary in two and three dimensions when compared with the variational method. Despite of this discrepancy between our two analytical approaches, their theoretical formulations suggest that both converge to the true result once higher hopping orders are taken into account.

For $d=1$ the quantum phase boundary of the Bose-Hubbard model is more complicated as it is a Kosterlitz-Thouless type of phase transition \cite{sachdev,kosterlitz}. This non-analytic behavior is reproduced in Fig.~\ref{fig1} quite well
by both the precise Density-Matrix-Renormalization Group results and the strong-coupling expansion 
(see also Ref.~\cite{monien3}). However, our two analytical methods cannot deal with this. Our variational method yields, at least, a continuous phase boundary, but our field-theoretical method leads to a finite interval of the chemical potential where no real solution for the phase boundary exists. This finding is insofar consistent as both methods are expected to be applicable only for small values of the hopping parameter $t$.

In Fig.~\ref{fig2} we can clearly observe how the phase boundaries of our two methods approach the mean-field phase boundary as the dimension $d$ of the system increases which is consistent with the fact that at large dimensions the mean-field theory becomes exact \cite{sachdev}. Such an agreement with the mean-field theory for $d \rightarrow \infty$ can also be checked directly in Eqs.~(\ref{boundary1}), 
(\ref{f.order}), and (\ref{roots}) by using the explicit results for $\alpha_2^{(0)}$, $\alpha_2^{(1)}$, and $\alpha_2^{(2)}$ from Eqs.~(\ref{formulas}). 

Finally, we observe that so far our two methods yield a phase boundary for $d=3$ dimensions which turns out to be analytical at the lobe tip. This finding is consistent with the theory of critical phenomena as a quantum phase transition in $d$ spatial dimensions belongs effectively to the universality class of a standard phase transition in $d+1$ dimensions \cite{fischer,sachdev,monien}. In contrast to that, the strong-coupling expansion of Ref. \cite{monien} leads in each order to a pronounced artificial cusp at the lobe tip. At present, it remains open to resolve the analytical structure of lobe tips via QMC simulations \cite{montecarlo3}. This is certainly demanding as the lobe tip is the most sensible region of the Mott lobe with respect to finite-size scalings. We expect that also at higher orders our two methods will show an analytical lobe tip. This would make them then ideal tools for dealing with experimental situations where the MI-SF phase boundary is crossed at a fixed particle number per site.

\section{Conclusions and Outlook}

In this work we have presented two alternative analytical methods which improve substantially the quantitative results 
for the MI-SF quantum phase boundary for dimensions larger than one as shown in Fig.~\ref{fig1} and Table \ref{table}.
For the first Mott lobe our results are in the immediate vicinity of recent
high-precision Monte-Carlo simulations. Therefore, it would be interesting to investigate with future Monte-Carlo simulations whether the accuracy of our results increases or decreases for higher Mott lobes.

We expect that both methods converge to one and the same result if higher orders in the hopping energy would be taken into account. However, as can be seen in Table \ref{table}, at least up to the second order in the hooping parameter $t$, the field-theoretical method turns out to converge faster than the variational method. Another advantage of the second method is that the order parameter $\psi$ is, by definition, the square root of the condensate density, while for the first method the physical interpretation of $\psi$ remains unclear. 

In addition, we remark that both theories in Sections III and IV have been formulated so general that they are also applicable for finite temperatures. Thus, it should soon be possible to improve the finite-temperature mean-field MI-SF phase boundary which was recently derived in Refs. \cite{bru,bousante,krutinsky}. 

Both approaches presented here for a homogeneous optical lattice can be extended in a straight-forward way to the experimental situation where an additional harmonic confinement potential is superimposed to the periodic potential. Within a Thomas-Fermi approximation the overall harmonic potential is taken into account by introducing a local chemical potential. Thus, the particle density distribution of such an inhomogeneous system follows from cutting the Mott lobes of a homogeneous quantum phase diagram with a horizontal line. This yields  to a wedding cake structure which consists of alternating concentric insulating and superfluid layers \cite{gerbierpra,gerbier,FW06}. A precise description of the respective layer widths is of fundamental importance in view of a quantitative analysis of time-of-flight data \cite{hoffmann}.

Finally, we conclude with the observation that our present approaches are so far restricted to the Mott-insulating regime. In order to extend them to the superfluid regime, where we have a non-vanishing order parameter, would necessitate to determine higher order terms in the Landau expansion. In addition, to have access also to local statistical quantities as, for instance, correlations functions it is indispensable to develop a Ginzburg-Landau expansion where the spatially and temporally homogeneous order parameter is generalized to an order parameter field \cite{barry}.

\begin{acknowledgments}
We cordially thank Barry Bradlyn, Henrik Enoksen, Robert Graham, Alexander Hoffmann, Konstantin Krutitsky, 
Flavio Nogueira,  and Matthias Ohliger for 
stimulating discussions. 
Furthermore, we acknowledge the financial support from the German Academic Exchange Service (DAAD) and from the German Research Foundation within the Collaborative Research Center SBF/TR 12 Symmetries and Universality in Mesoscopic Systems.
\end{acknowledgments}

\appendix

\section{\label{appA} Linked-Cluster Theorem}

Different versions of the linked cluster expansion have so far been applied to both classical and quantum lattices systems.
This section is dedicated to applying the linked-cluster theorem \cite{gelfand} to the Bose-Hubbard Hamiltonian with currents. Our first step is to introduce the more general Hamiltonian:
\begin{equation}
\label{Hmod}
\hat{H}_{\rm BH} = -\sum_{\langle i, j\rangle} t_{\langle i, j\rangle} \hat{a}^{\dag}_{i}\hat{a}_{j} +
\sum_{i}\left(\frac{U}{2}\hat{a}^{\dag}_{i}\hat{a}^{\dag}_{i}\hat{a}_{i}\hat{a}_{i} -
\mu\hat{a}^{\dag}_{i}\hat{a}_{i}\right) + \sum_i \left( J^{\ast}_i \hat{a}_i + J_i \hat{a}_i^{\dagger}\right).
\end{equation}
The difference between the above Hamiltonian and the Hamiltonian (\ref{hamiltonian2}) resides in the more general coefficients $t_{\langle i, j\rangle}$, $J^{\ast}_i$, and $J_i$. Now the hopping parameter $t$ can have different values for each pair of nearest neighbor lattice sites $\langle i, j \rangle$ and the sources can have different values for each lattice site.

  Now imagine a given extensive quantity $G$ associated with the system described by (\ref{Hmod}). Such a quantity can be expanded in a power series of the indices $\{t_{\langle i, j\rangle}\}$:

\begin{equation}
\label{expansion}
 Q(\{ t_{\langle i, j\rangle} \})= \sum_{\{n_{\langle i, j\rangle}\}}q\{n_{\langle i, j\rangle}\} \prod_{\langle i, j\rangle} t_{\langle i, j\rangle}^{n_{\langle i, j\rangle}} .
\end{equation}
Here $\{ t_{\langle i, j\rangle}\}$ denotes the set of all coefficients associated with the nearest neighbor pairs and $\{ n_{\langle i, j\rangle} \}$ is a set of discrete indices also associated with the nearest neighbor pairs with each $n_{\langle i, j\rangle}$ running over the nonnegative integers. Without loss of generality, we can assume that $Q = 0$ if $t_{\langle i, j\rangle}=0$, which is equivalent to assume $q_{\{ 0, 0, \cdots \}}=0$.

A cluster is defined as any nonempty set of neighbor pairs $t_{\langle i, j\rangle}$. With this definition (\ref{expansion}) can be written as 
\begin{equation}
\label{expansion2}
 Q(\{ t_{\langle i, j\rangle} \})= \sum_C W(C) ,
\end{equation}
where the sum runs over all possible clusters $C$. The cluster weight $W(C)$ contains all terms in (\ref{expansion}) which have at least one power of $t_{\langle i, j\rangle}$ for all $\langle i, j\rangle$ contained in $C$ and no powers of any other $t_{\langle i, j\rangle}$. The {\it linked cluster theorem} states that the weight $W(C_d)$ associated with a disconnected cluster $C_d$ is zero and, therefore, only connected clusters must be taken into account in (\ref{expansion2}). Here a disconnected cluster represents any cluster which is just a union of disjoint nonempty subclusters. In order to verify the {\it linked cluster theorem} we can suppose that only the coefficients  $t_{\langle i, j\rangle}$ associated with the pairs $\langle i, j\rangle$ contained in two disjoint nonempty clusters $C_1$ and $C_2$ are nonzero and all other $t_{\langle i, j\rangle}$ are zero. The additivity of $Q$ then implies that
\begin{equation}
\label{verifiation}
 Q(C_1 \cup C_2) = Q(C_1) + Q(C_2) .
\end{equation}
It means that in (\ref{expansion}) there is no term which includes $t_{\langle i, j\rangle}$ for all $\langle i, j\rangle$ contained in $C_1 \cup C_2$, therefore we have $W(C_1 \cup C_2) =0$. 

Although in (\ref{Hmod}) a different $t_{\langle i, j\rangle}$ is used for each pair $\langle i, j\rangle$, in our applications we are interested in the special case where $t_{\langle i, j\rangle}= t$ for all $\langle i, j\rangle$. In this case clusters related to each other by symmetries such as translations, reflections, and rotations give the same contribution for $Q$. Each set of equivalent clusters can be represented by a graph $G$ and the lattice constant $L(G)$ is the number of topologically equivalent clusters represented by $G$ per lattice site. Using these definitions Eq.~(\ref{expansion2}) reduces to
\begin{equation}
\label{expansion3}
 Q(t)= N_{\rm s}\sum_G L(G) W(G) \, ,
\end{equation}
where $N_{\rm s}$ denotes the number of lattice sites.

\section{\label{appB} Diagrammatic Cluster Representation}
According to Appendix \ref{appA} a given cluster $C$ is associated with a weight $W(C)$ which is given by the terms in 
Eq.~(\ref{expansion}) which have the product of coefficients $t_{\langle i, j\rangle}$ corresponding to all pairs in $C$, and without coefficients corresponding to pairs not contained in $C$. Each of these terms contributing to a $W(C)$ can be further expanded in a power series of the currents $J_i$ and $J^{\ast}_i$. If we set now constant values to $t_{\langle i, j\rangle}$, $J_i$, and $J^{\ast}_i$ and consider $G$ as being the correction to the grand-canonical free energy due to $t$, $J$, and $J^{\ast}$, we obtain the power series (\ref{free.energy}) and (\ref{coefficients}). 

As we can see in (\ref{Hmod}) each coefficient $t_{\langle i, j\rangle}$ is associated with an annihilation operator $\hat{a}_j$ and a creation operator $\hat{a}^{\dagger}_i$, while each current $J_i$ and $J_i^{\ast}$ is associated with a creation and annihilation operator $\hat{a}^{\dagger}_i$ and $\hat{a}_i$, respectively. Therefore, a set of diagrams composed of $n$ internal lines and $2p$ external lines linked to points, which correspond to one and the same cluster, can be associated to each coefficient $\alpha_{2p}^{(n)}$. With this each coefficient $\alpha_{2p}^{(n)}$ can be written as a sum of diagrams where each diagram is multiplied by its lattice number. With this we obtain, for example, for $n=0$ and $n=1$ the diagrams (\ref{aI}) and (\ref{aII}).

%Here $2d$ is the lattice number associated with the diagram in (\ref{p2}).
An interesting property of the diagram (\ref{aII}) is that it is one-particle reducible, and therefore, similarly to the usual Feynman diagrams used in quantum field theory, it factorizes into its one-particle irreducible contributions \cite{kleinert2,zinn-justin}. Thus, the diagrams (\ref{aI}) and (\ref{aII}) are related via

\begin{equation}
\label{b1}
\parbox{20mm}{
\begin{fmfgraph*}(60,20)
  \fmfleft{i1}
  \fmfright{o1}
  \fmf{fermion,label.side=left}{i1,v1}
  \fmfv{decor.shape=circle, decor.filled=1, decor.size=.2h}{v1}
  \fmf{fermion, label.side=left}{v1,v2}
  \fmfv{decor.shape=circle, decor.filled=1, decor.size=.2h}{v2}
  \fmf{fermion, label.side=left}{v2,o1} 
  \end{fmfgraph*}} \;= \left( \parbox{16mm}{
\begin{fmfgraph*}(40,20)
  \fmfleft{i1}
  \fmfright{o1}
  \fmf{fermion,label.side=left}{i1,v1}
  \fmfv{decor.shape=circle, decor.filled=1, decor.size=.2h}{v1}
  \fmf{fermion, label.side=left}{v1,o1}
  \end{fmfgraph*}} \right)^2 ,
\end{equation}
which yields in total
\begin{equation}
 \label{star}
 \alpha^{(1)}_2 = (2d) (\alpha^{(0)}_2)^2 .
\end{equation}

\section{\label{appC}Diagrammatic Notation for Time-Independent Perturbation Theory}

In this section a diagrammatic version of the usual Rayleigh-Schr\"{o}dinger perturbation theory is presented.
Using this approach the diagrams mentioned in the previous section are further decomposed into simpler contributions which are then easily computed.
 
In the conventional time-independent perturbation theory \cite{sakurai} the aim is to solve the eigenvalue problem

\begin{equation}
\label{schroedinger}
\hat{H} \vert\Psi_n \rangle = E_n \vert\Psi_n\rangle ,
\end{equation}
where $\hat{H} = \hat{H_0} + \lambda \hat{V}$, in power series of the smallness parameter $\lambda$:

\begin{eqnarray}
\vert \Psi_n \rangle &=& \sum_{i=0}^{\infty} \lambda^i \vert \Psi_{n}^{(i)} \rangle, \label{evector}\\
E_n &=&  \sum_{i=0}^{\infty} \lambda^i E_{n}^{(i)}. \label{evalue}
\end{eqnarray}
Solving the Schr\"{o}dinger equation (\ref{schroedinger}) with (\ref{evector}) and (\ref{evalue}) 
leads to the following recursion formula for $i>0$:

\begin{eqnarray}
E_{n}^{(i)} &=&  \langle\Psi_{n}^{(0)} \vert \hat{V} \vert \Psi_{n}^{(i-1)} \rangle, \nonumber \\
\vert \Psi_{n}^{(i)} \rangle &=& \sum_{m\neq n} \vert \Psi_{m}^{(0)} \rangle \frac{\langle\Psi_{m}^{(0)} \vert \hat{V} \vert \Psi_{n}^{(i-1)} \rangle}{E_{n}^{(0)}-E_{m}^{(0)}} - \sum_{j=1}^{i} E_{n}^{(j)} \sum_{m\neq n} \vert \Psi_{m}^{(0)} \rangle \frac{\langle\Psi_{m}^{(0)} \vert \Psi_{n}^{(i-j)} \rangle}{E_{n}^{(0)}-E_{m}^{(0)}}. \label{p.formula}
\end{eqnarray}
The first three terms of the eigenvalue expansion read explicitly:

\begin{eqnarray}
\label{explicit}
E_{n}^{(1)} &=&  \langle\Psi_{n}^{(0)} \vert \hat{V} \vert \Psi_{n}^{(0)} \rangle, \\
E_{n}^{(2)} &=&  \sum_{m\neq n}\frac{1}{E_{n}^{(0)}-E_{m}^{(0)}} \langle\Psi_{n}^{(0)}\vert\hat{V}\vert \Psi_{m}^{(0)}\rangle \langle\Psi_{m}^{(0)}\vert\hat{V}\vert \Psi_{n}^{(0)}\rangle ,  \label{p.formula2} \\
E_{n}^{(3)} &=& \sum_{m_1\neq n}\sum_{m_2\neq n}\frac{1}{E_{n}^{(0)}-E_{m_1}^{(0)}}\frac{1}{E_{n}^{(0)}-E_{m_2}^{(0)}}
\langle\Psi_{n}^{(0)}\vert\hat{V}\vert \Psi_{m_2}^{(0)}\rangle\langle\Psi_{m_2}^{(0)}\vert\hat{V}\vert \Psi_{m_1}^{(0)}\rangle\langle\Psi_{m_1}^{(0)}\vert\hat{V}\vert \Psi_{n}^{(0)}\rangle \nonumber \\ &&
- \sum_{m_1\neq n}\frac{1}{(E_{n}^{(0)}-E_{m_1}^{(0)})^2}\langle\Psi_{n}^{(0)}\vert\hat{V}\vert \Psi_{n}^{(0)}\rangle\langle\Psi_{n}^{(0)}\vert\hat{V}\vert \Psi_{m_1}^{(0)}\rangle\langle\Psi_{m_1}^{(0)}\vert\hat{V}\vert \Psi_{n}^{(0)}\rangle.\label{explicit2}
\end{eqnarray}
The results in Eqs.~(\ref{explicit})--(\ref{explicit2}) can be graphically represented by introducing a diagrammatic notation:

\begin{eqnarray}
E_{n}^{(1)} &=& \parbox{12mm}{
  \begin{fmfgraph}(30,20)
  \fmfleft{i1}
  \fmfright{o1}
  \fmf{plain}{i1,v1}	
  \fmfv{decor.shape=circle, decor.filled=1, decor.size=.3h}{v1}
  \fmf{plain}{v1,o1}
  \end{fmfgraph}} , \label{gr1}\\
E_{n}^{(2)} &=& \parbox{12mm}{\begin{fmfgraph*}(40,20)
  \fmfleft{i1}
  \fmfright{o1}
  \fmf{plain}{i1,v1}
  \fmfv{decor.shape=circle, decor.filled=1, decor.size=.3h}{v1}
  \fmf{plain}{v1,v2}
  \fmfv{decor.shape=circle, decor.filled=1, decor.size=.3h}{v2}
  \fmf{plain}{v2,o1}
  \end{fmfgraph*}} \;\;, \label{gr2} \\
E_{n}^{(3)} &=& \parbox{15mm}{\begin{fmfgraph*}(40,20)
  \fmfleft{i1}
  \fmfright{o1}
  \fmf{plain}{i1,v1}
  \fmfv{decor.shape=circle, decor.filled=1, decor.size=.3h}{v1}
  \fmf{plain}{v1,v2}
  \fmfv{decor.shape=circle, decor.filled=1, decor.size=.3h}{v2}
  \fmf{plain}{v2,v3}
  \fmfv{decor.shape=circle, decor.filled=1, decor.size=.3h}{v3}
  \fmf{plain}{v3,o1}
  \end{fmfgraph*}} - \parbox{20mm}{\begin{fmfgraph*}(40,20)
  \fmfleft{i1,i2}
  \fmfright{o1,o2}
  \fmf{plain}{i1,v1}
  \fmfv{decor.shape=circle, decor.filled=1, decor.size=.3h}{v1}
  \fmf{plain, left=.4, tension=.3}{v1,v2}
  \fmf{plain, right=.4, tension=.3}{v1,v2}
  \fmfv{decor.shape=circle, decor.filled=1, decor.size=.3h}{v2}
  \fmf{plain}{v2,o1}
  \fmf{plain}{i2,v3}
  \fmfv{decor.shape=circle, decor.filled=1, decor.size=.3h}{v3}
  \fmf{plain}{v3,o2}
  \end{fmfgraph*}} .\label{gr3}
\end{eqnarray}
Here the diagrams in Eqs.~(\ref{gr1})--(\ref{gr3}) must be read from right to left and are interpreted according to the following rules:

\begin{itemize}

\item Each dot represents an interaction $\hat{V}$. Therefore, the number of dots contained in a diagram defines the respective power of $\lambda$ this diagram corresponds to.
\item The internal lines, which appear between two consecutive dots, are associated with the factor $\sum_{m\neq n}\frac{1}{\left( E_{n}^{(0)} - E_{m}^{(0)}\right)^q}\vert\Psi_{m}^{(0)}\rangle\langle\Psi_{m}^{(0)}\vert $, where $q$ denotes the number of lines linking two given consecutive dots.
\item If we have more than one line between two consecutive dots, each extra line is associated with an extra disconnected part of the diagram. Note that each diagram has a prefactor $(-1)^{s-1}$, where $s$ is the total number of its disconnected parts.
\item The external lines are associated with the unperturbed bra and ket $\langle\Psi_{n}^{(0)}\vert\cdots\vert \Psi_{n}^{(0)}\rangle$.
\end{itemize}
For example, the diagram (\ref{gr1}) has just one dot and the external lines, the diagram (\ref{gr2}) consists of two dots, one internal line, and the external lines. Equation (\ref{gr3}) contains two diagrams, where the first one has three dots with two internal lines and the external lines. The second one represents a disconnected diagram where the lower part has two consecutive dots with two internal
lines between them. The extra internal line is associated with the upper disconnected part which has just one dot without internal lines. Using the prescription above we can easily reconstruct the formulas (\ref{explicit})--(\ref{explicit2}).

Following this prescription the $i$-th term $E_{n}^{(i)}$ of the perturbative expansion of the energy $E_n$ can be constructed by summing all the diagrams with $i$ dots and multiplying each diagram by a factor $(-1)^{s-1}$, where $s$ is the number of disconnected parts of the diagram.

Thus, the 4th order term of the energy reads diagrammatically

\begin{equation}
\label{ggr}
E_{n}^{(4)}=\parbox{21mm}{\begin{fmfgraph*}(60,20)
  \fmfleft{i1}
  \fmfright{o1}
  \fmf{plain}{i1,v1}
  \fmfv{decor.shape=circle, decor.filled=1, decor.size=.3h}{v1}
  \fmf{plain}{v1,v2}
  \fmfv{decor.shape=circle, decor.filled=1, decor.size=.3h}{v2}
  \fmf{plain}{v2,v3}
  \fmfv{decor.shape=circle, decor.filled=1, decor.size=.3h}{v3}
  \fmf{plain}{v3,v4}
  \fmfv{decor.shape=circle, decor.filled=1, decor.size=.3h}{v4}
  \fmf{plain}{v4,o1}
  \end{fmfgraph*}} -  \parbox{21mm}{\begin{fmfgraph*}(60,20)
  \fmfleft{i1,i2}
  \fmfright{o1,o2}
  \fmf{plain}{i1,v1}
  \fmfv{decor.shape=circle, decor.filled=1, decor.size=.3h}{v1}
  \fmf{plain, left=0.4, tension=.3}{v1,v2}
  \fmf{plain, right=0.4, tension=.3}{v1,v2}
  \fmfv{decor.shape=circle, decor.filled=1, decor.size=.3h}{v2}
  \fmf{plain}{v2,v3}
  \fmfv{decor.shape=circle, decor.filled=1, decor.size=.3h}{v3}
  \fmf{plain}{v3,o1}
  \fmf{plain}{i2,v4}
  \fmfv{decor.shape=circle, decor.filled=1, decor.size=.3h}{v4}
  \fmf{plain}{v4,o2}
  \end{fmfgraph*}} - \parbox{21mm}{\begin{fmfgraph*}(60,20)
  \fmfleft{i1,i2}
  \fmfright{o1,o2}
  \fmf{plain}{i1,v1}
  \fmfv{decor.shape=circle, decor.filled=1, decor.size=.3h}{v1}
  \fmf{plain}{v1,v2}
  \fmfv{decor.shape=circle, decor.filled=1, decor.size=.3h}{v2}
  \fmf{plain, left=.4, tension=.3}{v2,v3}
  \fmf{plain, right=.4, tension=.3}{v2,v3}
  \fmfv{decor.shape=circle, decor.filled=1, decor.size=.3h}{v3}
  \fmf{plain}{v3,o1}
  \fmf{plain}{i2,v4}
  \fmfv{decor.shape=circle, decor.filled=1, decor.size=.3h}{v4}
  \fmf{plain}{v4,o2}
  \end{fmfgraph*}}- \parbox{20mm}{\begin{fmfgraph*}(50,20)
  \fmfleft{i1,i2}
  \fmfright{o1,o2}
  \fmf{plain}{i1,v1}
  \fmfv{decor.shape=circle, decor.filled=1, decor.size=.3h}{v1}
  \fmf{plain, left=.4, tension=.3}{v1,v2}
  \fmf{plain, right=.4, tension=.3}{v1,v2}
  \fmfv{decor.shape=circle, decor.filled=1, decor.size=.3h}{v2}
  \fmf{plain}{v2,o1}
  \fmf{plain}{i2,v3}
  \fmfv{decor.shape=circle, decor.filled=1, decor.size=.3h}{v3}
  \fmf{plain}{v3,v4}
  \fmfv{decor.shape=circle, decor.filled=1, decor.size=.3h}{v4}
  \fmf{plain}{v4,o2}
  \end{fmfgraph*}}+ \parbox{20mm}{\begin{fmfgraph*}(50,30)
  \fmfleft{i1,i2,i3}
  \fmfright{o1,o2,o3}
  \fmf{plain}{i1,v1}
  \fmfv{decor.shape=circle, decor.filled=1, decor.size=.2h}{v1}
  \fmf{plain, left=.4, tension=.2}{v1,v2}
  \fmf{plain, right=.4, tension=.2}{v1,v2}
  \fmf{plain, tension=.2}{v1,v2}
  \fmfv{decor.shape=circle, decor.filled=1, decor.size=.2h}{v2}
  \fmf{plain}{v2,o1}
  \fmf{plain}{i2,v3}
  \fmfv{decor.shape=circle, decor.filled=1, decor.size=.2h}{v3}
  \fmf{plain}{v3,o2}
  \fmf{plain}{i3,v4}
  \fmfv{decor.shape=circle, decor.filled=1, decor.size=.2h}{v4}
  \fmf{plain}{v4,o3}
  \end{fmfgraph*}}.
\end{equation}
Using the prescription to write down the explicit formula for $E_n^{(4)}$ we have:
\begin{eqnarray}
   E_{n}^{\left( 4\right)} &=& \sum_{m_1 \neq n}\sum_{m_2 \neq n}\sum_{m_3 \neq n}
 \langle\Psi_{n}^{(0)}\vert\hat{V}
 \frac{\vert\Psi_{m_1}^{(0)}\rangle\langle\Psi_{m_1}^{(0)}\vert}{E_{m_1}^{(0)}-E_{n}^{(0)}}\hat{V}
 \frac{\vert\Psi_{m_2}^{(0)}\rangle\langle\Psi_{m_2}^{(0)}\vert}{E_{m_2}^{(0)}-E_{n}^{(0)}}\hat{V}
 \frac{\vert\Psi_{m_3}^{(0)}\rangle\langle\Psi_{m_3}^{(0)}\vert}{E_{m_2}^{(0)}-E_{n}^{(0)}}\hat{V}
 \vert\Psi_{n}^{(0)}\rangle \nonumber \\ &&
   -\sum_{m_1 \neq n}\sum_{m_2 \neq n}\langle\Psi_{n}^{(0)}\vert\hat{V}\vert\Psi_{n}^{(0)}\rangle
 \langle\Psi_{n}^{(0)}\vert\hat{V}
 \frac{\vert\Psi_{m_1}^{(0)}\rangle\langle\Psi_{m_1}^{(0)}\vert}{\left(
 E_{m_1}^{(0)}-E_{n}^{(0)}\right)^2}\hat{V}
 \frac{\vert\Psi_{m_2}^{(0)}\rangle\langle\Psi_{m_2}^{(0)}\vert}{E_{m_2}^{(0)}-E_{n}^{(0)}}\hat{V}
 \vert\Psi_{n}^{(0)}\rangle \nonumber \\ &&
   -\sum_{m_1 \neq n}\sum_{m_2 \neq n}\langle\Psi_{n}^{(0)}\vert\hat{V}\vert\Psi_{n}^{(0)}\rangle
 \langle\Psi_{n}^{(0)}\vert\hat{V}
 \frac{\vert\Psi_{m_1}^{(0)}\rangle\langle\Psi_{m_1}^{(0)}\vert}{E_{m_1}^{(0)}-E_{n}^{(0)}}\hat{V}
 \frac{\vert\Psi_{m_2}^{(0)}\rangle\langle\Psi_{m_2}^{(0)}\vert}{\left(
 E_{m_2}^{(0)}-E_{n}^{(0)}\right)^2}\hat{V}
 \vert\Psi_{n}^{(0)}\rangle \\ &&
   -\sum_{m_1 \neq n}\sum_{m_2 \neq n}\langle\Psi_{n}^{(0)}\vert\hat{V}
 \frac{\vert\Psi_{m_1}^{(0)}\rangle\langle\Psi_{m_1}^{(0)}\vert}{\left(
 E_{m_1}^{(0)}-E_{n}^{(0)}\right)^2}\hat{V}
 \vert\Psi_{n}^{(0)}\rangle
 \langle\Psi_{n}^{(0)}\vert\hat{V}
 \frac{\vert\Psi_{m_2}^{(0)}\rangle\langle\Psi_{m_2}^{(0)}\vert}{E_{m_2}^{(0)}-E_{n}^{(0)}}\hat{V}
 \vert\Psi_{n}^{(0)}\rangle \nonumber \\ &&
   +\sum_{m_1 \neq n}\langle\Psi_{n}^{(0)}\vert\hat{V}\vert\Psi_{n}^{(0)}\rangle
 \langle\Psi_{n}^{(0)}\vert\hat{V}\vert\Psi_{n}^{(0)}\rangle
 \langle\Psi_{n}^{(0)}\vert\hat{V}
 \frac{\vert\Psi_{m_1}^{(0)}\rangle\langle\Psi_{m_1}^{(0)}\vert}{\left(
 E_{m_1}^{(0)}-E_{n}^{(0)}\right)^3}\hat{V}
 \vert\Psi_{n}^{(0)}\rangle . \nonumber
\end{eqnarray}

Now we can construct diagrammatically any term of the perturbative series without using the recursion formulas (\ref{p.formula}) explicitly. Note that we read off from (\ref{p.formula}) a recursion formula for the total multiplicity $M_k$ of all diagrams in a given perturbative order:
\begin{equation}
 M_{k+1}=M_k + \sum_{l=1}^{k-1}M_{k-l+1}M_l .
\end{equation}
Starting from the initial value $M_1=1$ corresponding to (\ref{gr1}), its recursive solution gives $M_2=1$, $M_3=2$, and $M_4=5$, in agreement with  (\ref{gr2})--(\ref{ggr}) and yields furthermore $M_5=14$ and $M_6=42$ for the next two orders.

 This approach can also be used if more interactions are present. For example, if we 
 have $\hat{H}=\hat{H_0}+\lambda\hat{V}+\sigma\hat{W}$, then we can introduce the two
 vertices

\begin{eqnarray}
  \hat{V}&\rightarrow& \parbox{12mm}{\begin{fmfgraph*}(10,20)
  \fmfleft{i1}
  \fmfright{o1}
  \fmf{plain}{i1,v1}
  \fmfv{decor.shape=circle, decor.filled=1, decor.size=.3h,label.angle=90,label=$1$}{v1}
  \fmf{plain}{v1,o1}
  \end{fmfgraph*}} , \\ \nonumber \\
  \hat{W}&\rightarrow& \parbox{12mm}{\begin{fmfgraph*}(10,20)
  \fmfleft{i1}
  \fmfright{o1}
  \fmf{plain}{i1,v1}
  \fmfv{decor.shape=circle, decor.filled=1, decor.size=.3h,label.angle=90,label=$2$}{v1}
  \fmf{plain}{v1,o1}
  \end{fmfgraph*}}.
\end{eqnarray}
With this we obtain for the perturbative expansion
\begin{eqnarray}
E_n &=& E_{n}^{(0)}+\lambda \parbox{12mm}{\begin{fmfgraph*}(30,20)
  \fmfleft{i1}
  \fmfright{o1}
  \fmf{plain}{i1,v1}
  \fmfv{decor.shape=circle, decor.filled=1, decor.size=.3h,label.angle=90,label=$1$}{v1}
  \fmf{plain}{v1,o1}
  \end{fmfgraph*}}+\sigma \parbox{12mm}{\begin{fmfgraph*}(30,20)
  \fmfleft{i1}
  \fmfright{o1}
  \fmf{plain}{i1,v1}
  \fmfv{decor.shape=circle, decor.filled=1, decor.size=.3h,label.angle=90,label=$2$}{v1}
  \fmf{plain}{v1,o1}
  \end{fmfgraph*}}+ \lambda \sigma \left( \parbox{14mm}{\begin{fmfgraph*}(40,20)
  \fmfleft{i1}
  \fmfright{o1}
  \fmf{plain}{i1,v1}
  \fmfv{decor.shape=circle, decor.filled=1, decor.size=.3h,label.angle=90,label=$1$}{v1}
  \fmf{plain}{v1,v2}
  \fmfv{decor.shape=circle, decor.filled=1, decor.size=.3h,label.angle=90,label=$2$}{v2}
  \fmf{plain}{v2,o1}
  \end{fmfgraph*}} +\parbox{14mm}{\begin{fmfgraph*}(40,20)
  \fmfleft{i1}
  \fmfright{o1}
  \fmf{plain}{i1,v1}
  \fmfv{decor.shape=circle, decor.filled=1, decor.size=.3h,label.angle=90,label=$2$}{v1}
  \fmf{plain}{v1,v2}
  \fmfv{decor.shape=circle, decor.filled=1, decor.size=.3h,label.angle=90,label=$1$}{v2}
  \fmf{plain}{v2,o1}
  \end{fmfgraph*}} \right) + \lambda^2 \parbox{14mm}{\begin{fmfgraph*}(40,20)
  \fmfleft{i1}
  \fmfright{o1}
  \fmf{plain}{i1,v1}
  \fmfv{decor.shape=circle, decor.filled=1, decor.size=.3h,label.angle=90,label=$1$}{v1}
  \fmf{plain}{v1,v2}
  \fmfv{decor.shape=circle, decor.filled=1, decor.size=.3h,label.angle=90,label=$1$}{v2}
  \fmf{plain}{v2,o1} 
  \end{fmfgraph*}} \nonumber \\ \nonumber \\ && + \sigma^2 \parbox{14mm}{\begin{fmfgraph*}(40,20)
  \fmfleft{i1}
  \fmfright{o1}
  \fmf{plain}{i1,v1}
  \fmfv{decor.shape=circle, decor.filled=1, decor.size=.3h,label.angle=90,label=$2$}{v1}
  \fmf{plain}{v1,v2}
  \fmfv{decor.shape=circle, decor.filled=1, decor.size=.3h,label.angle=90,label=$2$}{v2}
  \fmf{plain}{v2,o1} 
  \end{fmfgraph*}} + \lambda^2 \sigma \bigg( \parbox{18mm}{\begin{fmfgraph*}(50,20)
  \fmfleft{i1}
  \fmfright{o1}
  \fmf{plain}{i1,v1}
  \fmfv{decor.shape=circle, decor.filled=1, decor.size=.3h,label.angle=90,label=$1$}{v1}
  \fmf{plain}{v1,v2}
  \fmfv{decor.shape=circle, decor.filled=1, decor.size=.3h,label.angle=90,label=$1$}{v2}
  \fmf{plain}{v2,v3}
  \fmfv{decor.shape=circle, decor.filled=1, decor.size=.3h,label.angle=90,label=$2$}{v3}
  \fmf{plain}{v3,o1}
  \end{fmfgraph*}} + \parbox{18mm}{\begin{fmfgraph*}(50,20)
  \fmfleft{i1}
  \fmfright{o1}
  \fmf{plain}{i1,v1}
  \fmfv{decor.shape=circle, decor.filled=1, decor.size=.3h,label.angle=90,label=$1$}{v1}
  \fmf{plain}{v1,v2}
  \fmfv{decor.shape=circle, decor.filled=1, decor.size=.3h,label.angle=90,label=$2$}{v2}
  \fmf{plain}{v2,v3}
  \fmfv{decor.shape=circle, decor.filled=1, decor.size=.3h,label.angle=90,label=$1$}{v3}
  \fmf{plain}{v3,o1}
  \end{fmfgraph*}} +\parbox{18mm}{\begin{fmfgraph*}(50,20)
  \fmfleft{i1}
  \fmfright{o1}
  \fmf{plain}{i1,v1}
  \fmfv{decor.shape=circle, decor.filled=1, decor.size=.3h,label.angle=90,label=$2$}{v1}
  \fmf{plain}{v1,v2}
  \fmfv{decor.shape=circle, decor.filled=1, decor.size=.3h,label.angle=90,label=$1$}{v2}
  \fmf{plain}{v2,v3}
  \fmfv{decor.shape=circle, decor.filled=1, decor.size=.3h,label.angle=90,label=$1$}{v3}
  \fmf{plain}{v3,o1}
  \end{fmfgraph*}}  - \parbox{20mm}{\begin{fmfgraph*}(50,10)
  \fmfleft{i1,i2}
  \fmfright{o1,o2}
  \fmf{plain}{i1,v1}
  \fmfv{decor.shape=circle, decor.filled=1, decor.size=.5h,label.angle=-90,label=$2$}{v1}
  \fmf{plain, left=.4, tension=.3}{v1,v2}
  \fmf{plain, right=.4, tension=.3}{v1,v2}
  \fmfv{decor.shape=circle, decor.filled=1, decor.size=.5h,label.angle=-90,label=$1$}{v2}
  \fmf{plain}{v2,o1}
  \fmf{plain}{i2,v3}
  \fmfv{decor.shape=circle, decor.filled=1, decor.size=.5h,label.angle=90,label=$1$}{v3}
  \fmf{plain}{v3,o2}
  \end{fmfgraph*}} \nonumber  \\ \nonumber \\ &&  - \parbox{20mm}{\begin{fmfgraph*}(50,10)
  \fmfleft{i1,i2}
  \fmfright{o1,o2}
  \fmf{plain}{i1,v1}
  \fmfv{decor.shape=circle, decor.filled=1, decor.size=.5h,label.angle=-90,label=$1$}{v1}
  \fmf{plain,left=.4, tension=.3}{v1,v2}
  \fmf{plain, right=.4, tension=.3}{v1,v2}
  \fmfv{decor.shape=circle, decor.filled=1, decor.size=.5h,label.angle=-90,label=$2$}{v2}
  \fmf{plain}{v2,o1}
  \fmf{plain}{i2,v3}
  \fmfv{decor.shape=circle, decor.filled=1, decor.size=.5h,label.angle=90,label=$1$}{v3}
  \fmf{plain}{v3,o2}
  \end{fmfgraph*}} - \parbox{20mm}{\begin{fmfgraph*}(50,10)
  \fmfleft{i1,i2}
  \fmfright{o1,o2}
  \fmf{plain}{i1,v1}
  \fmfv{decor.shape=circle, decor.filled=1, decor.size=.5h,label.angle=-90,label=$1$}{v1}
  \fmf{plain,left=.4, tension=.3}{v1,v2}
  \fmf{plain, right=.4, tension=.3}{v1,v2}
  \fmfv{decor.shape=circle, decor.filled=1, decor.size=.5h,label.angle=-90,label=$1$}{v2}
  \fmf{plain}{v2,o1}
  \fmf{plain}{i2,v3}
  \fmfv{decor.shape=circle, decor.filled=1, decor.size=.5h,label.angle=90,label=$2$}{v3}
  \fmf{plain}{v3,o2}
  \end{fmfgraph*}}\bigg) + \cdots .
\end{eqnarray}

\section{\label{appD} Explicit Evaluation of Diagrams}

The purpose of this section is to evaluate the `arrow' diagrams presented in Appendix \ref{appB} by using the `line' diagrams presented in Appendix \ref{appC}.

As we saw in Appendix \ref{appB} each line in an arrow diagram represents a process which is associated with the operators $\hat{a}_i$, $\hat{a}^{\dagger}_i$, or $\hat{a}^{\dagger}_j \hat{a}_i$ with $i$ and $j$ being nearest neighbor lattice sites. In the line diagrams each of these processes are represented by a point. Let us take as an example the following arrow diagram:

\begin{equation}
\label{nl}
 \parbox{40mm}{
\begin{fmfgraph*}(80,20)
  \fmfleft{i1}
  \fmfright{o1}
  \fmf{fermion,label.side=left,label=$1$}{i1,v1}
  \fmf{fermion,label.side=left, label=$3$}{v1,v2}
  \fmf{fermion,label.side=left, label=$2$}{v2,o1}
  \fmfv{decor.shape=circle, decor.filled=1, decor.size=.2h,label.angle=-90,label=$i$}{v1}
  \fmfv{decor.shape=circle, decor.filled=1, decor.size=.2h,label.angle=-90,label=$j$}{v2}
  \end{fmfgraph*}} .
\end{equation}
Here we labeled the processes with the numbers $1$, $3$, and $2$ which, according to (\ref{nl}), are associated with the operators $\hat{a}^{\dagger}_i$, $\hat{a}^{\dagger}_j \hat{a}_i$, and $\hat{a}_j$, respectively. This diagram belongs to a cluster which consists only of two neighbouring sites $i$ and $j$. Therefore, we define an effective Hamiltonian which takes into account only the relevant sites $i$ and $j$, as well as the relevant interaction terms corresponding to $\hat{a}^{\dagger}_i$, $\hat{a}^{\dagger}_j \hat{a}_i$, and $\hat{a}_j$ as follows: 

\begin{eqnarray}
\label{heff}
\hat{H}_{\rm eff} &=& \hat{H}_i + \hat{H}_j + \hat{V}_1 + \hat{V}_2+ \hat{V}_3.
\end{eqnarray}
Here the Hamiltonians $\hat{H}_i$ and $\hat{H}_j$ are defined according to (\ref{lcl}) and have the same set of eigenvalues given by
\begin{equation}
\label{eps}
\epsilon_n = \frac{U}{2} (n^2 -n) -\mu n 
\end{equation}
and the interactions read
\begin{eqnarray}
 \hat{V}_1 = J \hat{a}^{\dagger}_i\,, \hspace{1cm}
 \hat{V}_2 = J^{\ast} \hat{a}_j\,, \hspace{1cm}
 \hat{V}_3 = -t \hat{a}^{\dagger}_i \hat{a}_j \,. 
\end{eqnarray}
The value of (\ref{nl}) is given by that term in the expansion of the ground-state energy corresponding to 
(\ref{heff}) which is proportional to $-t J^{\ast} J$. According to our discussion in Appendix \ref{appC}, this term can be split using line diagrams. Therefore the diagram (\ref{nl}) can be written as follows: 
\begin{equation}
 \parbox{20mm}{
\begin{fmfgraph*}(60,20)
  \fmfleft{i1}
  \fmfright{o1}
  \fmf{fermion,label.side=left, label=$1$}{i1,v1}
  \fmf{fermion,label.side=left, label=$3$}{v1,v2}
  \fmf{fermion,label.side=left, label=$2$}{v2,o1}
  \fmfv{decor.shape=circle, decor.filled=1, decor.size=.2h}{v1}
  \fmfv{decor.shape=circle, decor.filled=1, decor.size=.2h}{v2}
  \end{fmfgraph*}} \;=  \parbox{20mm}{
\begin{fmfgraph*}(60,20)
  \fmfleft{i1}
  \fmfright{o1}
  \fmf{plain}{i1,v1}
  \fmf{plain}{v1,v2}
  \fmf{plain}{v2,v3}
  \fmf{plain}{v3,o1}
  \fmfv{decor.shape=circle, decor.filled=1, decor.size=.3h,label.angle=90,label=$1$}{v1}
  \fmfv{decor.shape=circle, decor.filled=1, decor.size=.3h,label.angle=90,label=$2$}{v2}
  \fmfv{decor.shape=circle, decor.filled=1, decor.size=.3h,label.angle=90,label=$3$}{v3}
  \end{fmfgraph*}} \; + \parbox{20mm}{
\begin{fmfgraph*}(60,20)
  \fmfleft{i1}
  \fmfright{o1}
  \fmf{plain}{i1,v1}
  \fmf{plain}{v1,v2}
  \fmf{plain}{v2,v3}
  \fmf{plain}{v3,o1}
  \fmfv{decor.shape=circle, decor.filled=1, decor.size=.3h,label.angle=90,label=$1$}{v1}
  \fmfv{decor.shape=circle, decor.filled=1, decor.size=.3h,label.angle=90,label=$3$}{v2}
  \fmfv{decor.shape=circle, decor.filled=1, decor.size=.3h,label.angle=90,label=$2$}{v3}
  \end{fmfgraph*}} \; + \text{more 4 permutations}.
\end{equation}
Each line diagram can now be evaluated using the rules presented in Appendix \ref{appB}. As an example we consider the line diagram

\begin{equation}
 g_{1,3,2}=  \parbox{40mm}{\begin{fmfgraph*}(60,20)
  \fmfleft{i1}
  \fmfright{o1}
  \fmf{plain}{i1,v1}
  \fmf{plain}{v1,v2}
  \fmf{plain}{v2,v3}
  \fmf{plain}{v3,o1}
  \fmfv{decor.shape=circle, decor.filled=1, decor.size=.3h,label.angle=90,label=$1$}{v1}
  \fmfv{decor.shape=circle, decor.filled=1, decor.size=.3h,label.angle=90,label=$3$}{v2}
  \fmfv{decor.shape=circle, decor.filled=1, decor.size=.3h,label.angle=90,label=$2$}{v3}
  \end{fmfgraph*}} ,
\end{equation}
which must be evaluated as follows:
\begin{itemize}
 \item Reading the diagram from right to left we find first the dot number two which is
       associated with the operator $\hat{a}_j$. It annihilates one particle at site $j$ generating the
       factors $\sqrt{n}$ and $1/(\epsilon_n -\epsilon_{n-1})$ and leaving $n-1$ at this site:
       \begin{equation}
       g_{1,3,2} = \sqrt{n}\frac{1}{\epsilon_n -\epsilon_{n-1}} \cdots .
       \end{equation}	
 \item The next dot is number three which is associated with $ \hat{a}^{\dagger}_j \hat{a}_i $. Thus a particle is created at site $j$ while another one is annihilated at site $i$. This leads to two factors
       $\sqrt[2]{n}$ and the factor $1/(\epsilon_n -\epsilon_{n-1})$:
       \begin{equation}
       g_{1,3,2} = \sqrt{n}\frac{1}{\epsilon_n -\epsilon_{n-1}} n \frac{1}{\epsilon_n -\epsilon_{n-1}}\cdots .
       \end{equation}
 \item The last dot is number one which is associated with $\hat{a}^{\dagger}_i$. It creates the last 
       particle at site $i$ and yields the additional factor $\sqrt{n}$. Thus, we obtain at the end:
       \begin{equation}
         g_{1,3,2} = n^2 \frac{1}{(\epsilon_n -\epsilon_{n-1})^2}.
       \end{equation}
\end{itemize}

Now we present a complete list of all arrow diagrams up to the second hopping order which have been evaluated using line diagrams. In order to simplify the notation we will use the definitions
\begin{eqnarray}
\label{def1}
\lambda^{+p} &=& \epsilon_n - \epsilon_{n+p} , \qquad 
\lambda^{-p} = \epsilon_n - \epsilon_{n-p}, \\
\lambda^{+} &=& \lambda^{+1},  \qquad\qquad\;\;\;
\lambda^{-} = \lambda^{-1} \label{def2} .
\end{eqnarray}

%Following is a complete list of all arrow diagrams evaluated by using point diagrams, %where the diagrams (\ref{l1}), (\ref{l2}), and (\ref{l3}) leads directly to %(\ref{alpha0}), (\ref{alpha1}), and (\ref{alpha3}) respectively:
We start with the calculation of the arrow diagram (\ref{aI}) which decomposes into two line diagrams:

\begin{equation}
\label{d11}
\parbox{20mm}{
\begin{fmfgraph*}(60,20)
  \fmfleft{i1}
  \fmfright{o1}
  \fmf{fermion,label.side=left, label=$1$}{i1,v1}
  \fmf{fermion,label.side=left, label=$2$}{v1,o1}
  \fmfv{decor.shape=circle, decor.filled=1, decor.size=.2h}{v1}
  \end{fmfgraph*}} \;= 
\parbox{20mm}{\begin{fmfgraph*}(60,20)
  \fmfleft{i1}
  \fmfright{o1}
  \fmf{plain}{i1,v1}
  \fmf{plain}{v1,v2}
  \fmf{plain}{v2,o1}
  \fmfv{decor.shape=circle, decor.filled=1, decor.size=.3h,label.angle=90,label=$1$}{v1}
  \fmfv{decor.shape=circle, decor.filled=1, decor.size=.3h,label.angle=90,label=$2$}{v2}
  \end{fmfgraph*}} \;+
\parbox{20mm}{\begin{fmfgraph*}(60,20)
  \fmfleft{i1}
  \fmfright{o1}
  \fmf{plain}{i1,v1}
  \fmf{plain}{v1,v2}
  \fmf{plain}{v2,o1}
  \fmfv{decor.shape=circle, decor.filled=1, decor.size=.3h,label.angle=90,label=$2$}{v1}
  \fmfv{decor.shape=circle, decor.filled=1, decor.size=.3h,label.angle=90,label=$1$}{v2}
  \end{fmfgraph*}} \qquad .
\end{equation}
Both contributions in (\ref{d11}) lead to the following expressions
\begin{eqnarray}
\parbox{20mm}{\begin{fmfgraph*}(60,20)
  \fmfleft{i1}
  \fmfright{o1}
  \fmf{plain}{i1,v1}
  \fmf{plain}{v1,v2}
  \fmf{plain}{v2,o1}
  \fmfv{decor.shape=circle, decor.filled=1, decor.size=.3h,label.angle=90,label=$1$}{v1}
  \fmfv{decor.shape=circle, decor.filled=1, decor.size=.3h,label.angle=90,label=$2$}{v2}
  \end{fmfgraph*}} \;&=& (n+1) \frac{1}{\lambda^{+}} , \\
\parbox{20mm}{\begin{fmfgraph*}(60,20)
  \fmfleft{i1}
  \fmfright{o1}
  \fmf{plain}{i1,v1}
  \fmf{plain}{v1,v2}
  \fmf{plain}{v2,o1}
  \fmfv{decor.shape=circle, decor.filled=1, decor.size=.3h,label.angle=90,label=$2$}{v1}
  \fmfv{decor.shape=circle, decor.filled=1, decor.size=.3h,label.angle=90,label=$1$}{v2}
  \end{fmfgraph*}} &\;=& n \frac{1}{\lambda^{-}} ,
\end{eqnarray}
thus yielding in total the result
\begin{equation}
\parbox{20mm}{\begin{fmfgraph*}(60,20)
  \fmfleft{i1}
  \fmfright{o1}
  \fmf{fermion,label.side=left}{i1,v1}
  \fmf{fermion,label.side=left}{v1,o1}
  \fmfv{decor.shape=circle, decor.filled=1, decor.size=.2h}{v1}
  \end{fmfgraph*}} \;= \frac{n}{\lambda^{-}} + \frac{n+1}{\lambda^{+}} . \label{l1}
\end{equation}
Combining (\ref{def1}) and (\ref{def2}) with (\ref{eps}) and (\ref{l1}) thus results in (\ref{alpha0})
with the abbreviation $b=\mu / U$.

Correspondingly, the arrow diagram in (\ref{aII}) factorizes according to (\ref{b1}), so we obtain with (\ref{l1})
\begin{equation}
\parbox{20mm}{
\begin{fmfgraph*}(60,20)
  \fmfleft{i1}
  \fmfright{o1}
  \fmf{fermion,label.side=left}{i1,v1}
  \fmfv{decor.shape=circle, decor.filled=1, decor.size=.2h}{v1}
  \fmf{fermion, label.side=left}{v1,v2}
  \fmfv{decor.shape=circle, decor.filled=1, decor.size=.2h}{v2}
  \fmf{fermion, label.side=left}{v2,o1} 
  \end{fmfgraph*}} \;= \left( \parbox{16mm}{
\begin{fmfgraph*}(40,20)
  \fmfleft{i1}
  \fmfright{o1}
  \fmf{fermion,label.side=left}{i1,v1}
  \fmfv{decor.shape=circle, decor.filled=1, decor.size=.2h}{v1}
  \fmf{fermion, label.side=left}{v1,o1}
  \end{fmfgraph*}} \right)^2 \;= \left( \frac{n}{\lambda^{-}} + \frac{n+1}{\lambda^{+}} \right)^2 ,\label{l2}
\end{equation}
which leads to the result (\ref{alpha1}).

The first arrow diagram in (\ref{aIII}) factorizes in a similar way, yielding
\begin{equation}
\parbox{20mm}{
\begin{fmfgraph*}(60,20)
  \fmfleft{i1}
  \fmfright{o1}
  \fmf{fermion,label.side=left}{i1,v1}
  \fmfv{decor.shape=circle, decor.filled=1, decor.size=.2h}{v1}
  \fmf{fermion, label.side=left}{v1,v2}
  \fmfv{decor.shape=circle, decor.filled=1, decor.size=.2h}{v2}
  \fmf{fermion, label.side=left}{v2,v3} 
  \fmfv{decor.shape=circle, decor.filled=1, decor.size=.2h}{v3}
  \fmf{fermion, label.side=left}{v3,o1} 
  \end{fmfgraph*}} \;= \left( \parbox{16mm}{
\begin{fmfgraph*}(40,20)
  \fmfleft{i1}
  \fmfright{o1}
  \fmf{fermion,label.side=left}{i1,v1}
  \fmfv{decor.shape=circle, decor.filled=1, decor.size=.2h}{v1}
  \fmf{fermion, label.side=left}{v1,o1}
  \end{fmfgraph*}} \right)^3 \;= \left( \frac{n}{\lambda^{-}} + \frac{n+1}{\lambda^{+}} \right)^3 .
\end{equation}

Now we return to the second arrow diagram in (\ref{aIII}) which decomposes in line diagrams according to:

\begin{equation}
\parbox{20mm}{
\begin{fmfgraph*}(60,40)
  \fmfleft{i1}
  \fmfright{o1}
  \fmfbottom{p1}
  \fmftop{u1}
  \fmf{fermion,label.side=right,label=$2$}{i1,v1}
  \fmfv{decor.shape=circle, decor.filled=1, decor.size=.1h}{v1}
  \fmf{fermion,tension=0,left=0.4,label.side=left,label=$4$}{v1,u1}
  \fmfv{decor.shape=circle, decor.filled=1, decor.size=.1h}{u1}
  \fmf{fermion,tension=0,left=0.4,label.side=left,label=$3$}{u1,v1}
  \fmf{fermion,label.side=right,label=$1$}{v1,o1} 
  \end{fmfgraph*}} \; = \parbox{20mm}{\begin{fmfgraph*}(60,20)
  \fmfleft{i1}
  \fmfright{o1}
  \fmf{plain}{i1,v1}
  \fmf{plain}{v1,v2}
  \fmf{plain}{v2,v3}
  \fmf{plain}{v3,v4}
  \fmf{plain}{v4,o1}
  \fmfv{decor.shape=circle, decor.filled=1, decor.size=.3h}{v1}
  \fmfv{decor.shape=circle, decor.filled=1, decor.size=.3h}{v2}
  \fmfv{decor.shape=circle, decor.filled=1, decor.size=.3h}{v3}
  \fmfv{decor.shape=circle, decor.filled=1, decor.size=.3h}{v4}
  \end{fmfgraph*}} \; - \parbox{20mm}{\begin{fmfgraph*}(60,20)
  \fmfbottom{i1,o1}
  \fmftop{i2,o2}
  \fmf{plain, tension=2}{i1,v1}
  \fmf{plain, right=0.3}{v1,v2}
  \fmf{plain, left=0.3}{v1,v2}
  \fmf{plain, tension=2}{v2,o1}
  \fmf{plain}{i2,v3}
  \fmf{plain}{v3,v4}
  \fmf{plain}{v4,o2}
  \fmfv{decor.shape=circle, decor.filled=1, decor.size=.3h}{v1}
  \fmfv{decor.shape=circle, decor.filled=1, decor.size=.3h}{v2}
  \fmfv{decor.shape=circle, decor.filled=1, decor.size=.3h}{v3}
  \fmfv{decor.shape=circle, decor.filled=1, decor.size=.3h}{v4}
  \end{fmfgraph*}} \qquad. \label{l3}
\end{equation}
Here  the non-numbered dot diagrams represent a sum over all numbered diagrams with the respective topology. Their explicit evaluation yields

\begin{eqnarray}
\label{cc1}
 \parbox{20mm}{\begin{fmfgraph*}(60,20)
  \fmfleft{i1}
  \fmfright{o1}
  \fmf{plain}{i1,v1}
  \fmf{plain}{v1,v2}
  \fmf{plain}{v2,v3}
  \fmf{plain}{v3,v4}
  \fmf{plain}{v4,o1}
  \fmfv{decor.shape=circle, decor.filled=1, decor.size=.3h,label.angle=90,label=$1$}{v1}
  \fmfv{decor.shape=circle, decor.filled=1, decor.size=.3h,label.angle=90,label=$3$}{v2}
  \fmfv{decor.shape=circle, decor.filled=1, decor.size=.3h,label.angle=90,label=$2$}{v3}
  \fmfv{decor.shape=circle, decor.filled=1, decor.size=.3h,label.angle=90,label=$4$}{v4}
  \end{fmfgraph*}} \;&=&  \frac{n (n+1)^2}{(\lambda^{+})^2 (\lambda^{+} + \lambda^{-})} , \\
 \parbox{20mm}{\begin{fmfgraph*}(60,20)
  \fmfleft{i1}
  \fmfright{o1}
  \fmf{plain}{i1,v1}
  \fmf{plain}{v1,v2}
  \fmf{plain}{v2,v3}
  \fmf{plain}{v3,v4}
  \fmf{plain}{v4,o1}
  \fmfv{decor.shape=circle, decor.filled=1, decor.size=.3h,label.angle=90,label=$1$}{v1}
  \fmfv{decor.shape=circle, decor.filled=1, decor.size=.3h,label.angle=90,label=$4$}{v2}
  \fmfv{decor.shape=circle, decor.filled=1, decor.size=.3h,label.angle=90,label=$2$}{v3}
  \fmfv{decor.shape=circle, decor.filled=1, decor.size=.3h,label.angle=90,label=$3$}{v4}
  \end{fmfgraph*}} \;&=& \frac{n (n+1) (n+2)}{\lambda^{+} (\lambda^{+2} + \lambda^{-}) (\lambda^{+} + \lambda^{-})} , \\
 \parbox{20mm}{\begin{fmfgraph*}(60,20)
  \fmfleft{i1}
  \fmfright{o1}
  \fmf{plain}{i1,v1}
  \fmf{plain}{v1,v2}
  \fmf{plain}{v2,v3}
  \fmf{plain}{v3,v4}
  \fmf{plain}{v4,o1}
  \fmfv{decor.shape=circle, decor.filled=1, decor.size=.3h,label.angle=90,label=$1$}{v1}
  \fmfv{decor.shape=circle, decor.filled=1, decor.size=.3h,label.angle=90,label=$3$}{v2}
  \fmfv{decor.shape=circle, decor.filled=1, decor.size=.3h,label.angle=90,label=$4$}{v3}
  \fmfv{decor.shape=circle, decor.filled=1, decor.size=.3h,label.angle=90,label=$2$}{v4}
  \end{fmfgraph*}} \;&=& \frac{(n+1)^3}{(\lambda^{+})^3} , \\
 \parbox{20mm}{\begin{fmfgraph*}(60,20)
  \fmfleft{i1}
  \fmfright{o1}
  \fmf{plain}{i1,v1}
  \fmf{plain}{v1,v2}
  \fmf{plain}{v2,v3}
  \fmf{plain}{v3,v4}
  \fmf{plain}{v4,o1}
  \fmfv{decor.shape=circle, decor.filled=1, decor.size=.3h,label.angle=90,label=$1$}{v1}
  \fmfv{decor.shape=circle, decor.filled=1, decor.size=.3h,label.angle=90,label=$4$}{v2}
  \fmfv{decor.shape=circle, decor.filled=1, decor.size=.3h,label.angle=90,label=$3$}{v3}
  \fmfv{decor.shape=circle, decor.filled=1, decor.size=.3h,label.angle=90,label=$2$}{v4}
  \end{fmfgraph*}} \;&=& \frac{n (n+1) (n+2)}{(\lambda^{+})^2 (\lambda^{+2} + \lambda^{-})} , \\
 \parbox{20mm}{\begin{fmfgraph*}(60,20)
  \fmfleft{i1}
  \fmfright{o1}
  \fmf{plain}{i1,v1}
  \fmf{plain}{v1,v2}
  \fmf{plain}{v2,v3}
  \fmf{plain}{v3,v4}
  \fmf{plain}{v4,o1}
  \fmfv{decor.shape=circle, decor.filled=1, decor.size=.3h,label.angle=90,label=$3$}{v1}
  \fmfv{decor.shape=circle, decor.filled=1, decor.size=.3h,label.angle=90,label=$1$}{v2}
  \fmfv{decor.shape=circle, decor.filled=1, decor.size=.3h,label.angle=90,label=$2$}{v3}
  \fmfv{decor.shape=circle, decor.filled=1, decor.size=.3h,label.angle=90,label=$4$}{v4}
  \end{fmfgraph*}} \;&=& \frac{n^2 (n+1)}{\lambda^{+} (\lambda^{+} + \lambda^{-})^2} , \\
 \parbox{20mm}{\begin{fmfgraph*}(60,20)
  \fmfleft{i1}
  \fmfright{o1}
  \fmf{plain}{i1,v1}
  \fmf{plain}{v1,v2}
  \fmf{plain}{v2,v3}
  \fmf{plain}{v3,v4}
  \fmf{plain}{v4,o1}
  \fmfv{decor.shape=circle, decor.filled=1, decor.size=.3h,label.angle=90,label=$4$}{v1}
  \fmfv{decor.shape=circle, decor.filled=1, decor.size=.3h,label.angle=90,label=$1$}{v2}
  \fmfv{decor.shape=circle, decor.filled=1, decor.size=.3h,label.angle=90,label=$2$}{v3}
  \fmfv{decor.shape=circle, decor.filled=1, decor.size=.3h,label.angle=90,label=$3$}{v4}
  \end{fmfgraph*}} \;&=& \frac{n (n+1) (n+2)}{(\lambda^{+2} + \lambda^{-}) (\lambda^{+} + \lambda^{-})^2} , \\
 \parbox{20mm}{\begin{fmfgraph*}(60,20)
  \fmfleft{i1}
  \fmfright{o1}
  \fmf{plain}{i1,v1}
  \fmf{plain}{v1,v2}
  \fmf{plain}{v2,v3}
  \fmf{plain}{v3,v4}
  \fmf{plain}{v4,o1}
  \fmfv{decor.shape=circle, decor.filled=1, decor.size=.3h,label.angle=90,label=$3$}{v1}
  \fmfv{decor.shape=circle, decor.filled=1, decor.size=.3h,label.angle=90,label=$1$}{v2}
  \fmfv{decor.shape=circle, decor.filled=1, decor.size=.3h,label.angle=90,label=$4$}{v3}
  \fmfv{decor.shape=circle, decor.filled=1, decor.size=.3h,label.angle=90,label=$2$}{v4}
  \end{fmfgraph*}} \;&=& \frac{n (n+1)^2}{(\lambda^{+})^2 (\lambda^{+} + \lambda^{-})} , \\
  \parbox{20mm}{\begin{fmfgraph*}(60,20)
  \fmfleft{i1}
  \fmfright{o1}
  \fmf{plain}{i1,v1}
  \fmf{plain}{v1,v2}
  \fmf{plain}{v2,v3}
  \fmf{plain}{v3,v4}
  \fmf{plain}{v4,o1}
  \fmfv{decor.shape=circle, decor.filled=1, decor.size=.3h,label.angle=90,label=$4$}{v1}
  \fmfv{decor.shape=circle, decor.filled=1, decor.size=.3h,label.angle=90,label=$1$}{v2}
  \fmfv{decor.shape=circle, decor.filled=1, decor.size=.3h,label.angle=90,label=$3$}{v3}
  \fmfv{decor.shape=circle, decor.filled=1, decor.size=.3h,label.angle=90,label=$2$}{v4}
  \end{fmfgraph*}} \;&=& \frac{n (n+1) (n+2)}{\lambda^{+} (\lambda^{+2} + \lambda^{-}) (\lambda^{+} + \lambda^{-})} , \\
 \parbox{20mm}{\begin{fmfgraph*}(60,20)
  \fmfleft{i1}
  \fmfright{o1}
  \fmf{plain}{i1,v1}
  \fmf{plain}{v1,v2}
  \fmf{plain}{v2,v3}
  \fmf{plain}{v3,v4}
  \fmf{plain}{v4,o1}
  \fmfv{decor.shape=circle, decor.filled=1, decor.size=.3h,label.angle=90,label=$2$}{v1}
  \fmfv{decor.shape=circle, decor.filled=1, decor.size=.3h,label.angle=90,label=$3$}{v2}
  \fmfv{decor.shape=circle, decor.filled=1, decor.size=.3h,label.angle=90,label=$1$}{v3}
  \fmfv{decor.shape=circle, decor.filled=1, decor.size=.3h,label.angle=90,label=$4$}{v4}
  \end{fmfgraph*}} \;&=& \frac{n (n-1) (n+1)}{\lambda^{-} (\lambda^{+} + \lambda^{-2}) (\lambda^{+} + \lambda^{-})} , \\
 \parbox{20mm}{\begin{fmfgraph*}(60,20)
  \fmfleft{i1}
  \fmfright{o1}
  \fmf{plain}{i1,v1}
  \fmf{plain}{v1,v2}
  \fmf{plain}{v2,v3}
  \fmf{plain}{v3,v4}
  \fmf{plain}{v4,o1}
  \fmfv{decor.shape=circle, decor.filled=1, decor.size=.3h,label.angle=90,label=$2$}{v1}
  \fmfv{decor.shape=circle, decor.filled=1, decor.size=.3h,label.angle=90,label=$4$}{v2}
  \fmfv{decor.shape=circle, decor.filled=1, decor.size=.3h,label.angle=90,label=$1$}{v3}
  \fmfv{decor.shape=circle, decor.filled=1, decor.size=.3h,label.angle=90,label=$3$}{v4}
  \end{fmfgraph*}} \;&=& \frac{n^2 (n+1)}{(\lambda^{-})^2 (\lambda^{+} + \lambda^{-})} , 
\end{eqnarray}
\begin{eqnarray}
 \parbox{20mm}{\begin{fmfgraph*}(60,20)
  \fmfleft{i1}
  \fmfright{o1}
  \fmf{plain}{i1,v1}
  \fmf{plain}{v1,v2}
  \fmf{plain}{v2,v3}
  \fmf{plain}{v3,v4}
  \fmf{plain}{v4,o1}
  \fmfv{decor.shape=circle, decor.filled=1, decor.size=.3h,label.angle=90,label=$3$}{v1}
  \fmfv{decor.shape=circle, decor.filled=1, decor.size=.3h,label.angle=90,label=$2$}{v2}
  \fmfv{decor.shape=circle, decor.filled=1, decor.size=.3h,label.angle=90,label=$1$}{v3}
  \fmfv{decor.shape=circle, decor.filled=1, decor.size=.3h,label.angle=90,label=$4$}{v4}
  \end{fmfgraph*}} \;&=& \frac{n (n-1) (n+1)}{(\lambda^{+} + \lambda^{-2}) (\lambda^{+} + \lambda^{-})^2} , \\
 \parbox{20mm}{\begin{fmfgraph*}(60,20)
  \fmfleft{i1}
  \fmfright{o1}
  \fmf{plain}{i1,v1}
  \fmf{plain}{v1,v2}
  \fmf{plain}{v2,v3}
  \fmf{plain}{v3,v4}
  \fmf{plain}{v4,o1}
  \fmfv{decor.shape=circle, decor.filled=1, decor.size=.3h,label.angle=90,label=$4$}{v1}
  \fmfv{decor.shape=circle, decor.filled=1, decor.size=.3h,label.angle=90,label=$2$}{v2}
  \fmfv{decor.shape=circle, decor.filled=1, decor.size=.3h,label.angle=90,label=$1$}{v3}
  \fmfv{decor.shape=circle, decor.filled=1, decor.size=.3h,label.angle=90,label=$3$}{v4}
  \end{fmfgraph*}} \;&=& \frac{n (n+1)^2}{\lambda^{-} (\lambda^{+} + \lambda^{-})^2} , \\
 \parbox{20mm}{\begin{fmfgraph*}(60,20)
  \fmfleft{i1}
  \fmfright{o1}
  \fmf{plain}{i1,v1}
  \fmf{plain}{v1,v2}
  \fmf{plain}{v2,v3}
  \fmf{plain}{v3,v4}
  \fmf{plain}{v4,o1}
  \fmfv{decor.shape=circle, decor.filled=1, decor.size=.3h,label.angle=90,label=$2$}{v1}
  \fmfv{decor.shape=circle, decor.filled=1, decor.size=.3h,label.angle=90,label=$3$}{v2}
  \fmfv{decor.shape=circle, decor.filled=1, decor.size=.3h,label.angle=90,label=$4$}{v3}
  \fmfv{decor.shape=circle, decor.filled=1, decor.size=.3h,label.angle=90,label=$1$}{v4}
  \end{fmfgraph*}} \;&=& \frac{n (n-1) (n+1)}{(\lambda^{-})^2 (\lambda^{+} + \lambda^{-2})} , \\
 \parbox{20mm}{\begin{fmfgraph*}(60,20)
  \fmfleft{i1}
  \fmfright{o1}
  \fmf{plain}{i1,v1}
  \fmf{plain}{v1,v2}
  \fmf{plain}{v2,v3}
  \fmf{plain}{v3,v4}
  \fmf{plain}{v4,o1}
  \fmfv{decor.shape=circle, decor.filled=1, decor.size=.3h,label.angle=90,label=$2$}{v1}
  \fmfv{decor.shape=circle, decor.filled=1, decor.size=.3h,label.angle=90,label=$4$}{v2}
  \fmfv{decor.shape=circle, decor.filled=1, decor.size=.3h,label.angle=90,label=$3$}{v3}
  \fmfv{decor.shape=circle, decor.filled=1, decor.size=.3h,label.angle=90,label=$1$}{v4}
  \end{fmfgraph*}} \;&=& \frac{n^3}{(\lambda^{-})^3} , \\
 \parbox{20mm}{\begin{fmfgraph*}(60,20)
  \fmfleft{i1}
  \fmfright{o1}
  \fmf{plain}{i1,v1}
  \fmf{plain}{v1,v2}
  \fmf{plain}{v2,v3}
  \fmf{plain}{v3,v4}
  \fmf{plain}{v4,o1}
  \fmfv{decor.shape=circle, decor.filled=1, decor.size=.3h,label.angle=90,label=$3$}{v1}
  \fmfv{decor.shape=circle, decor.filled=1, decor.size=.3h,label.angle=90,label=$2$}{v2}
  \fmfv{decor.shape=circle, decor.filled=1, decor.size=.3h,label.angle=90,label=$4$}{v3}
  \fmfv{decor.shape=circle, decor.filled=1, decor.size=.3h,label.angle=90,label=$1$}{v4}
  \end{fmfgraph*}} \;&=& \frac{n (n-1) (n+1)}{\lambda^{-} (\lambda^{+} + \lambda^{-}) (\lambda^{+} + \lambda^{-2})} , \\
 \parbox{20mm}{\begin{fmfgraph*}(60,20)
  \fmfleft{i1}
  \fmfright{o1}
  \fmf{plain}{i1,v1}
  \fmf{plain}{v1,v2}
  \fmf{plain}{v2,v3}
  \fmf{plain}{v3,v4}
  \fmf{plain}{v4,o1}
  \fmfv{decor.shape=circle, decor.filled=1, decor.size=.3h,label.angle=90,label=$4$}{v1}
  \fmfv{decor.shape=circle, decor.filled=1, decor.size=.3h,label.angle=90,label=$2$}{v2}
  \fmfv{decor.shape=circle, decor.filled=1, decor.size=.3h,label.angle=90,label=$3$}{v3}
  \fmfv{decor.shape=circle, decor.filled=1, decor.size=.3h,label.angle=90,label=$1$}{v4}
  \end{fmfgraph*}} \;&=& \frac{n^2 (n+1)}{(\lambda^{-})^2 (\lambda^{+} + \lambda^{-})} , \\
\parbox{20mm}{\begin{fmfgraph*}(60,20)
  \fmfleft{i1}
  \fmfright{o1}
  \fmf{plain}{i1,v1}
  \fmf{plain}{v1,v2}
  \fmf{plain}{v2,v3}
  \fmf{plain}{v3,v4}
  \fmf{plain}{v4,o1}
  \fmfv{decor.shape=circle, decor.filled=1, decor.size=.3h,label.angle=90,label=$4$}{v1}
  \fmfv{decor.shape=circle, decor.filled=1, decor.size=.3h,label.angle=90,label=$2$}{v2}
  \fmfv{decor.shape=circle, decor.filled=1, decor.size=.3h,label.angle=90,label=$3$}{v3}
  \fmfv{decor.shape=circle, decor.filled=1, decor.size=.3h,label.angle=90,label=$1$}{v4}
  \end{fmfgraph*}} \;&=& \frac{n^2 (n+1)}{(\lambda^{-})^2 (\lambda^{+} + \lambda^{-})},
\end{eqnarray} 
\begin{eqnarray}
\parbox{20mm}{\begin{fmfgraph*}(60,20)
  \fmfbottom{i1,o1}
  \fmftop{i2,o2}
  \fmf{plain, tension=2}{i1,v1}
  \fmf{plain, right=0.3}{v1,v2}
  \fmf{plain, left=0.3}{v1,v2}
  \fmf{plain, tension=2}{v2,o1}
  \fmf{plain}{i2,v3}
  \fmf{plain}{v3,v4}
  \fmf{plain}{v4,o2}
  \fmfv{decor.shape=circle, decor.filled=1, decor.size=.3h,label.angle=135,label=$1$}{v1}
  \fmfv{decor.shape=circle, decor.filled=1, decor.size=.3h,label.angle=45,label=$2$}{v2}
  \fmfv{decor.shape=circle, decor.filled=1, decor.size=.3h,label.angle=90,label=$3$}{v3}
  \fmfv{decor.shape=circle, decor.filled=1, decor.size=.3h,label.angle=90,label=$4$}{v4}
  \end{fmfgraph*}}\;&=&   \frac{n^2 (n+1)}{(\lambda^{-})^2 (\lambda^{+} + \lambda^{-})} , \\ \nonumber \\
\parbox{20mm}{\begin{fmfgraph*}(60,20)
  \fmfbottom{i1,o1}
  \fmftop{i2,o2}
  \fmf{plain, tension=2}{i1,v1}
  \fmf{plain, right=0.3}{v1,v2}
  \fmf{plain, left=0.3}{v1,v2}
  \fmf{plain, tension=2}{v2,o1}
  \fmf{plain}{i2,v3}
  \fmf{plain}{v3,v4}
  \fmf{plain}{v4,o2}
  \fmfv{decor.shape=circle, decor.filled=1, decor.size=.3h,label.angle=135,label=$1$}{v1}
  \fmfv{decor.shape=circle, decor.filled=1, decor.size=.3h,label.angle=45,label=$2$}{v2}
  \fmfv{decor.shape=circle, decor.filled=1, decor.size=.3h,label.angle=90,label=$4$}{v3}
  \fmfv{decor.shape=circle, decor.filled=1, decor.size=.3h,label.angle=90,label=$3$}{v4}
  \end{fmfgraph*}}\;&=&  \frac{n^2 (n+1)}{(\lambda^{-})^2 (\lambda^{+} + \lambda^{-})} , \\ \nonumber \\
\parbox{20mm}{\begin{fmfgraph*}(60,20)
  \fmfbottom{i1,o1}
  \fmftop{i2,o2}
  \fmf{plain, tension=2}{i1,v1}
  \fmf{plain, right=0.3}{v1,v2}
  \fmf{plain, left=0.3}{v1,v2}
  \fmf{plain, tension=2}{v2,o1}
  \fmf{plain}{i2,v3}
  \fmf{plain}{v3,v4}
  \fmf{plain}{v4,o2}
  \fmfv{decor.shape=circle, decor.filled=1, decor.size=.3h,label.angle=135,label=$2$}{v1}
  \fmfv{decor.shape=circle, decor.filled=1, decor.size=.3h,label.angle=45,label=$1$}{v2}
  \fmfv{decor.shape=circle, decor.filled=1, decor.size=.3h,label.angle=90,label=$3$}{v3}
  \fmfv{decor.shape=circle, decor.filled=1, decor.size=.3h,label.angle=90,label=$4$}{v4}
  \end{fmfgraph*}}\;&=&  \frac{n (n+1)^2}{(\lambda^{+})^2 (\lambda^{+} + \lambda^{-})} , \\ \nonumber \\
\parbox{20mm}{\begin{fmfgraph*}(60,20)
  \fmfbottom{i1,o1}
  \fmftop{i2,o2}
  \fmf{plain, tension=2}{i1,v1}
  \fmf{plain, right=0.3}{v1,v2}
  \fmf{plain, left=0.3}{v1,v2}
  \fmf{plain, tension=2}{v2,o1}
  \fmf{plain}{i2,v3}
  \fmf{plain}{v3,v4}
  \fmf{plain}{v4,o2}
  \fmfv{decor.shape=circle, decor.filled=1, decor.size=.3h,label.angle=135,label=$2$}{v1}
  \fmfv{decor.shape=circle, decor.filled=1, decor.size=.3h,label.angle=45,label=$1$}{v2}
  \fmfv{decor.shape=circle, decor.filled=1, decor.size=.3h,label.angle=90,label=$4$}{v3}
  \fmfv{decor.shape=circle, decor.filled=1, decor.size=.3h,label.angle=90,label=$3$}{v4}
  \end{fmfgraph*}}\;&=& \frac{n (n+1)^2}{(\lambda^{+})^2 (\lambda^{+} + \lambda^{-})}  , \\ \nonumber \\
\parbox{20mm}{\begin{fmfgraph*}(60,20)
  \fmfbottom{i1,o1}
  \fmftop{i2,o2}
  \fmf{plain, tension=2}{i1,v1}
  \fmf{plain, right=0.3}{v1,v2}
  \fmf{plain, left=0.3}{v1,v2}
  \fmf{plain, tension=2}{v2,o1}
  \fmf{plain}{i2,v3}
  \fmf{plain}{v3,v4}
  \fmf{plain}{v4,o2}
  \fmfv{decor.shape=circle, decor.filled=1, decor.size=.3h,label.angle=135,label=$3$}{v1}
  \fmfv{decor.shape=circle, decor.filled=1, decor.size=.3h,label.angle=45,label=$4$}{v2}
  \fmfv{decor.shape=circle, decor.filled=1, decor.size=.3h,label.angle=90,label=$1$}{v3}
  \fmfv{decor.shape=circle, decor.filled=1, decor.size=.3h,label.angle=90,label=$2$}{v4}
  \end{fmfgraph*}}\;&=& \frac{n^2 (n+1)}{\lambda^{-} (\lambda^{+} + \lambda^{-})^2} , \\ \nonumber \\
\parbox{20mm}{\begin{fmfgraph*}(60,20)
  \fmfbottom{i1,o1}
  \fmftop{i2,o2}
  \fmf{plain, tension=2}{i1,v1}
  \fmf{plain, right=0.3}{v1,v2}
  \fmf{plain, left=0.3}{v1,v2}
  \fmf{plain, tension=2}{v2,o1}
  \fmf{plain}{i2,v3}
  \fmf{plain}{v3,v4}
  \fmf{plain}{v4,o2}
  \fmfv{decor.shape=circle, decor.filled=1, decor.size=.3h,label.angle=135,label=$3$}{v1}
  \fmfv{decor.shape=circle, decor.filled=1, decor.size=.3h,label.angle=45,label=$4$}{v2}
  \fmfv{decor.shape=circle, decor.filled=1, decor.size=.3h,label.angle=90,label=$2$}{v3}
  \fmfv{decor.shape=circle, decor.filled=1, decor.size=.3h,label.angle=90,label=$1$}{v4}
  \end{fmfgraph*}}\;&=& \frac{n (n+1)^2}{\lambda^{+} (\lambda^{+} + \lambda^{-})^2} , \\ \nonumber \\
\parbox{20mm}{\begin{fmfgraph*}(60,20)
  \fmfbottom{i1,o1}
  \fmftop{i2,o2}
  \fmf{plain, tension=2}{i1,v1}
  \fmf{plain, right=0.3}{v1,v2}
  \fmf{plain, left=0.3}{v1,v2}
  \fmf{plain, tension=2}{v2,o1}
  \fmf{plain}{i2,v3}
  \fmf{plain}{v3,v4}
  \fmf{plain}{v4,o2}
  \fmfv{decor.shape=circle, decor.filled=1, decor.size=.3h,label.angle=135,label=$4$}{v1}
  \fmfv{decor.shape=circle, decor.filled=1, decor.size=.3h,label.angle=45,label=$3$}{v2}
  \fmfv{decor.shape=circle, decor.filled=1, decor.size=.3h,label.angle=90,label=$1$}{v3}
  \fmfv{decor.shape=circle, decor.filled=1, decor.size=.3h,label.angle=90,label=$2$}{v4}
  \end{fmfgraph*}}\;&=& \frac{n^2 (n+1)}{\lambda^{-} (\lambda^{+} + \lambda^{-})^2} ,\\ \nonumber \\
\parbox{20mm}{\begin{fmfgraph*}(60,20)
  \fmfbottom{i1,o1}
  \fmftop{i2,o2}
  \fmf{plain, tension=2}{i1,v1}
  \fmf{plain, right=0.3}{v1,v2}
  \fmf{plain, left=0.3}{v1,v2}
  \fmf{plain, tension=2}{v2,o1}
  \fmf{plain}{i2,v3}
  \fmf{plain}{v3,v4}
  \fmf{plain}{v4,o2}
  \fmfv{decor.shape=circle, decor.filled=1, decor.size=.3h,label.angle=135,label=$4$}{v1}
  \fmfv{decor.shape=circle, decor.filled=1, decor.size=.3h,label.angle=45,label=$3$}{v2}
  \fmfv{decor.shape=circle, decor.filled=1, decor.size=.3h,label.angle=90,label=$2$}{v3}
  \fmfv{decor.shape=circle, decor.filled=1, decor.size=.3h,label.angle=90,label=$1$}{v4}
  \end{fmfgraph*}}\;&=& \frac{n (n+1)^2}{\lambda^{+} (\lambda^{+} + \lambda^{-})^2} .\label{cc2}
\end{eqnarray}

Thus, combining (\ref{l2}) with (\ref{def1}), (\ref{def2}) and (\ref{l3})--(\ref{cc2}) finally yields (\ref{alpha2}).

\newpage %Just because of unusual number of tables stacked at end
\bibliography{bibliografia.bib}% Produces the bibliography via BibTeX.
\end{fmffile}
\end{document}